\documentclass[fleqn,usenatbib]{mnras}

\usepackage{float}
\usepackage{amssymb, amsmath, epsfig}
\usepackage{graphics,times,array,ctable,amsmath,amssymb,natbib}
\usepackage{threeparttable}

\newcommand{\bfk}{{\boldsymbol{k}}}

\newcommand{\bfx}{{\boldsymbol{x}}}

\newcommand{\nhat}{{\hat{\boldsymbol{n}}}}
\newcommand{\Mpc}{{\rm Mpc}}

\newcommand{\Msun}{{M_{\odot}}}

\newcommand{\HI}{H{\sc \ i}}

\def\apj{ApJ}
\def\apjl{ApJL}
\def\apjs{ApJS}

\def\mnras{MNRAS}
\def\physrep{physrep}

\def\araa{{Ann.\ Rev.\ Astron.\& Astrophys.\ }}
\def\aap{{A\&A}}
\def\prd{PRD}
\def\jcap{JCAP}

\title[Impact of UVB fluctuations on galaxy clustering]{Estimates for the impact of Ultraviolet Background fluctuations on galaxy clustering measurements}

\author[Upton Sanderbeck, Ir\v si\v c, McQuinn, \& Meiksin]{
Phoebe Upton Sanderbeck,$^{1,2}$\thanks{Email: phoebeu@ucr.edu} 
Vid Ir\v{s}i\v{c},$^2$ 
Matthew McQuinn,$^2$ 
and Avery Meiksin,$^3$
\\
$^1$Department of Physics \& Astronomy, University of California, Riverside\\
$^2$Astronomy Department, University of Washington\\
$^3$Scottish Universities Physics Alliance, Institute for Astronomy, University of Edinburgh
}

\pubyear{2018}
\begin{document}
\label{firstpage}
\pagerange{\pageref{firstpage}--\pageref{lastpage}}
\maketitle





\begin{abstract}         
Spatial fluctuations in ultraviolet backgrounds can subtly modulate the distribution of extragalactic sources, a potential signal and systematic for large-scale structure surveys.  While this modulation has been shown to be significant for 3D Ly$\alpha$ forest surveys, its relevance for other large-scale structure probes has been hardly explored, despite being the only astrophysical process that likely can affect clustering measurements on scales $\gtrsim$Mpc. We estimate that background fluctuations, modulating the amount of \HI, have a fractional effect of $(0.03-0.3)\times (k/[10^{-2} \Mpc^{-1}])^{-1}$ on the power spectrum of 21cm intensity maps at $z=1-3$.  We find a smaller effect for H$\alpha$ and Ly$\alpha$ intensity mapping surveys of $(0.001-0.1)\times (k/[10^{-2} \Mpc^{-1}])^{-1}$ and even smaller effect for more traditional surveys that correlate the positions of individual H$\alpha$ or Ly$\alpha$ emitters.  We also estimate the effect of backgrounds on low-redshift galaxy surveys in general based on a simple model in which background fluctuations modulate the rate halo gas cools, modulating star formation:  We estimate a maximum fractional effect on the power of $\sim 0.01 (k/[10^{-2} \Mpc^{-1}])^{-1}$ at $z=1$. We compare sizes of these imprints to cosmological parameter benchmarks for the next generation of redshift surveys: We find ionizing backgrounds could result in a bias on the squeezed triangle non-Gaussianity parameter $f_{\rm NL}$ that can be larger than unity for power spectrum measurements with a SPHEREx-like galaxy survey, and typical values of intensity bias. Marginalizing over a shape of the form $k^{-1} P_L$, where $P_L$ is the linear matter power spectrum, removes much of this bias at the cost of $\approx 40\%$ larger statistical errors. 
\end{abstract}


\section{Introduction}

Radiation backgrounds, especially of the ionizing sort, have the potential to impact structure in the Universe on larger scales than any other non-gravitational, astrophysical process. The photons in these backgrounds can travel significant fractions of the Hubble length \citep{meiksin93, haardt96,haardt12}, and, with their ability to ionize neutral hydrogen and alter the cooling rates of ions, they can modulate the properties of extragalactic sources \citep{benson02,hambrick09}. Despite this large-scale impact, the effect of background fluctuations on various large-scale tracers has only been investigated for 21cm intensity mapping surveys \citep{wyithe09}.  This paper attempts to estimate how ionizing background fluctuations impact post-reionization 21cm intensity mapping surveys, Ly$\alpha$ and H$\alpha$ emitter surveys, and galaxy surveys in general.  
  
Detecting the imprint of these fluctuations on large-scale structure (LSS) surveys would constrain the properties of the sources, such as the fraction of the ionizing background that owes to quasars versus galactic emissions \citep[e.g.,][]{2015ApJ...813L...8M}.  In addition, an imprint from background fluctuations in LSS surveys could complicate cosmological parameter constraints on, e.g., neutrino masses, the scalar spectral tilt plus its running, or primordial non-Gaussianity -- meeting motivated benchmarks for all of these parameters requires sub-percent precision on clustering measurements if not much better \citep{baldauf16}. Ionizing backgrounds have already been shown to be an important systematic for extracting cosmological parameters from 3D Ly$\alpha$ forest observations \citep{mcquinn11, pontzen14a, gontcho14, meiksin18}. We investigate here whether ionizing background fluctuations could also be relevant for other LSS measurements.\footnote{We note that the shapes imparted by ionizing background fluctuations are only captured by standard biasing expansions \citep[e.g.][]{2018PhR...733....1D} on scales larger than the mean free path (and produce non-local biasing terms).  Additionally, arguments that certain large-scale correlations must owe to primordial non-Gaussianity or equivalence principle violations, and cannot be mimicked by astrophysics \citep{2013NuPhB.873..514K, 2013JCAP...12..025C}, may be voided by ionizing backgrounds, as (unlike other astrophysical systematics) ionizing backgrounds propagate as far as the Horizon.}

A related effect that is not considered here is the effect of reionization on dwarf galaxies \citep{thoul96, gnedin00, noh14}, an effect that can potentially cascade up to more massive galaxies as the affected dwarfs merge and grow.  This could affect the clustering of galaxies, especially small galaxies at high redshifts \citep{2006ApJ...640....1B, pritchard07, 2007MNRAS.382..921W}.  The spectral shape of the effect of reionization is more difficult to predict than the shape imparted by the post-reionization background as considered here.  We suspect that the effects studied here are larger than the imprint of reionization, especially for the redshifts we consider ($z<3$).

This paper is organized as follows. Section~\ref{sec:ps} describes how the ionizing background affects linear scales in LSS measurements, and Section~\ref{sec:ionizingbackgrounds} outlines the formalism used to calculate the spectrum of ionizing background fluctuations and presents the fluctuation models this study employs. Section~\ref{sec:HI} estimates the impact of background fluctuations on post-reionization 21cm intensity mapping surveys. Section~\ref{sec:rr} presents similar estimates but for surveys that view galaxies by either their Ly$\alpha$ or H$\alpha$ emission. Section~\ref{sec:galsur} considers the effects of ionizing background inhomogeneities on galaxy surveys in general. Finally, Section~\ref{sec:nongaus} compares the requirements for constraining targeted cosmological parameters to benchmark values, a particularly relevant one (owing to its similar shape to that imparted by ionizing background fluctuations) being squeezed triangle primordial non-Gaussianity. 

\section{The impact of intensity fluctuations on galaxy clustering}
\label{sec:ps}

Metagalactic fluctuations angle-averaged ionizing intensity, $J$, have a component that traces the cosmic matter field with transfer function $T_J(k, w_\nu)$ and an effective source number density $n_J(k, w_\nu)$, where $w_\nu$ indicates some frequency-weighting of the radiation background and the $k$-dependence of both $T_J$ and $n_J$ account for the propagation of radiation. Thus, the linear power spectrum of intensity fluctuations is given by
\begin{equation}
P_J(k) = T_J(k, w_\nu)^2 P_{\delta_L}(k) + n_J(k, w_\nu)^{-1},
\end{equation}
where $P_{\delta_L}(k)$ is the linear matter power spectrum. In Section~\ref{sec:ionizingbackgrounds} we discuss how $T_J$ and $n_J$ are computed. For much larger wavenumbers than that set by the mean free path (and that set by the quasar lifetime for the `shot' $n_J^{-1}$ component), $T_J^2$ and $n_J^{-1}$ decrease as $k^{-2}$, resulting in the impact of ionizing backgrounds being largest at low wavenumbers.

Our concentration is not the nature of fluctuations, $P_J$ itself.  For $P_J$, we follow the calculations of \citet{meiksin18}, whose calculation we briefly summarize below.  Instead, our focus is the imprint that $J$ fluctuations have on LSS surveys.  When including intensity fluctuations, the standard expression for the linear power spectrum of some galaxy population is extended to
\begin{equation}
P_g(\bfk) = \left(b_g + b_J T_J + f \mu^2\right)^2 P_{\delta_L} + b_J^2  n_J^{-1},
\label{eqn:Pg}
\end{equation}
where $\mu\equiv \nhat \cdot \bfk$, $\nhat$ is the line-of-sight unit vector, $f\approx \Omega_m(z)^{0.6}$ and the associated term arises from redshift space distortions \citep{kaiser87}, and $b_g$ ($b_J$) are the linear density bias (intensity bias) of the particular galaxy population.  Equation~(\ref{eqn:Pg}) ignored that the $b_J$ coefficient could be different between the stochastic and density-tracing terms if the property that shapes galaxy observability is averaged over a longer time than the lifetime of sources, which would act to decrease the shot noise bias coefficient. Similarly, there should be a separate bias coefficient for the ionizing background at every previous redshift \citep{2018arXiv181202731C}.  Equation~(\ref{eqn:Pg}) made the simplification that the clustering is shaped by the background at the observed redshift, which should largely hold for the processes that couple the galaxy distribution to the radiation background considered here.

Since $b_J$ sets the magnitude of intensity fluctuations, much of the focus of this study is on estimating this bias for different types of galaxy survey. To preview our results, our estimates of $b_J$ from Sections~\ref{sec:HI}-\ref{sec:galsur} are summarized in Table~\ref{table:results}, in addition to our estimates for the fractional effect of intensity fluctuations on $P_g$. 

\begin{table*}
\caption{The impact of ionizing background fluctuations on different redshift surveys.  Here $b_J$ is our estimated response of the overdensity in the specified LSS observable to a fraction fluctuation in the background intensity.  We also list rough numbers for the fractional change in the specified LSS tracer's power spectrum from background fluctuations that use our models for $P_\Gamma$. \label{table:results}}

\begin{center}
\begin{threeparttable}
\begin{tabular}{l c c c c}
survey & redshifts & $b_J$  & fractional change in $P_g$ from $J$-fluctuations\\
\hline
21cm intensity  & $[1,2,3]$ & $[-0.25, -0.25, -0.25]$ &   $ [0.03,0.1,0.3] (k/[10^{-2} \Mpc^{-1}])^{-1}$ \\
H$\alpha$ or Ly$\alpha$ intensity & $[1,3]$ & $[0.008,0.03]$ & $ [0.001,0.1] (k/[10^{-2} \Mpc^{-1}])^{-1}$ \\
$L_\star$ galaxy surveys & $[0.5,2]$ & $[-0.05, -0.05]$\tnote{a} & $[0.002,0.03] (k/[10^{-2} \Mpc^{-1}])^{-1}$ \\
\end{tabular}
\begin{tablenotes}
\item[a] Like in the other rows, these values are used for calculating the third column; however, unlike in the other rows, there is a large range of possible values, with our estimates in \S~\ref{sec:galsur} finding $b_J=-(0.01-0.1)$.  These estimates should be taken with caution, assuming simplistically that the star formation rate of $L_\star$ galaxies is tied to the cooling rate of circumgalactic gas.
\end{tablenotes}
\end{threeparttable}
\end{center}
\end{table*}

\section{The linear theory of ionizing radiation backgrounds}  
\label{sec:ionizingbackgrounds}
On megaparsec scales and greater, fluctuations in the $z\lesssim5$ ionizing background are small and, hence, well described by perturbation theory. 
While the fluctuations are small, they may still be a relevant driver of inhomogeneity for large-scale structure (LSS) surveys. Aside from the Ly$\alpha$ forest, this source of inhomogeneity has not been previously considered. 

This section discusses our calculation of the linear power spectrum of intensity fluctuations, with our approach following the spirit of earlier calculations for intensity fluctuations \citep{meiksin04, croft04,mcdonald05, mcquinn11, pontzen14a,gontcho14, pontzen14b, 2017arXiv170602716S, meiksin18}. Our calculations especially follow the formalism presented in \citet{meiksin18}, which builds most closely off the approach in \citet{pontzen14a} but instead solves the fully time-dependent rather than the steady state solution.\footnote{Time dependence is even more important to include for our calculations than those in \citet{meiksin18}, which only considered $z>2$, as at low redshifts much of the background is comprised of photons that traveled a significant fraction of the Horizon.}  Future sections use these calculations to estimate the imprint on various LSS surveys.

To proceed, we linearize and then solve the cosmological radiative transfer equation, given by
\begin{equation}
\frac{\partial I_{\nu_0}}{\partial t} + 2 \frac{\dot a}{a}  I_{\nu_0}  + a^{-1} \,\nhat \cdot \nabla I_{\nu_0} = - \kappa_{\nu_0} I_{\nu_0}+  j_{\nu_0},
\end{equation}
where $\nu_0 \equiv \nu/(1+z)$, $I_\nu$ is the specific photon number intensity [photons~Hz$^{-1}$s$^{-1}$cm$^{-2}$sr$^{-1}$] -- noting that $J_\nu = (4\pi)^{-1} \int d\Omega I_\nu$ --, $a$ the cosmic scale factor, $\kappa_\nu$ and $j_\nu$ the absorption and emission coefficients, and we have set the speed of light to unity for simplicity.  

Linearizing this equation in the intensity fluctuation $\delta I_{\nu_0} \equiv I_{\nu_0} - \bar I_{\nu_0}$ and going to Fourier space yields
\begin{eqnarray}
 \dot {\widetilde{\delta I_{\nu_0}}} &  - i a^{-1} \nhat \cdot \bfk\; \widetilde{\delta I_{\nu_0}} +  \left(\bar \kappa_{\nu_0} +  2 H\right) \widetilde{\delta I_{\nu_0}} \nonumber \\
  & =   \tilde \beta_0 + \beta_{\delta} \tilde\delta + \beta_{\Gamma}  \tilde\delta_\Gamma; \label{eqn:linInu}\\
 \ &  \beta_{\delta}(\nu, t)  \equiv  \bar j_\nu \,c_{j, \delta} - \bar \kappa_{\nu} {\bar I_\nu} c_{\kappa, \delta}, \nonumber \\
 & \beta_{\Gamma}(\nu, t)  \equiv  \bar j_\nu c_{j, \Gamma}  - \bar \kappa_{\nu} {\bar I_{\nu}}  c_{\kappa, \Gamma}, \nonumber 
\end{eqnarray}
 where tildes indicate the field's Fourier dual and we have also expanded the absorption and emission coefficients to linear order in their overdensities in the density and in the photoionization rate, namely $\kappa_{\nu} = \bar \kappa_{\nu} \left(1 + c_{\kappa, \delta} \delta + c_{\kappa, \Gamma} \delta_\Gamma \right)$ and  $j_\nu = \bar j_\nu \left(1+ c_{j, \delta} \delta + c_{j, \Gamma} \delta_\Gamma \right)$, plus a stochastic field $\beta_0(\bfx)$ from uncorrelated small-scale structure.\footnote{The \HI\ photoionization rate is most relevant for any of the modulating effects of the ionizing background on galaxies (as essentially it modulates the opacity and recombination emission), a justification for only expanding the background in terms of $\delta \Gamma$ rather than the more general expansion in fluctuations in the specific intensity.  In detail, the expansion should also be in gas temperature; however, the temperature of photoionized gas correlates strongly with density, especially at the higher densities that are relevant.} The ionizing sources' bias should dominate $c_{j, \delta}$; we henceforth identify this coefficient with the sources' bias.

We further assume that quasars dominate the ultraviolet background, consistent with the findings of most background models at $z\lesssim 3$ \citep{1993ApJ...412...34M, haardt12, mcquinnHeII, uptonsanderbeck18}, although a dominant galactic contribution would only have an ${\cal O}(1)$ effect on our estimates for the imprint on $P_g$.\footnote{In the general case of both galaxies and quasars, the amplitude would instead be set by the bias of both populations weighted by their fractional contribution (because we find the density tracing term is dominant over the Poissonian, even for rare quasars).}  To model quasars, we use the $1~$Ry quasar emissivity of \citet{haardt12}, and we assume that their spectral index in specific intensity is $\alpha_Q = 1.7$.   For the quasar bias, we linearly interpolate between $c_{j, \delta} = \{0.5, 1.1, 1.5, 1.8, 2.2 \}$ at  $z=\{1.4, 1.8, 2.1, 2.6, 3.0\}$, numbers based on quasar SDSS clustering measurements \citep{ross09, 2017JCAP...07..017L}.  (Note that our ionizing source biases are denoted as $c_{j, X}$ to not confuse with the biases of the measured clustering signal $b_g$ and $b_j$; c.f.~Eqn.~\ref{eqn:Pg})   We further assume that $c_{j, \delta} \propto (1+z)$ to extrapolate to $z>3$.  While these bias measurements are for the $>L_*$ quasars observable with BOSS, we note that in many models for quasar clustering the luminosity dependence is weak \citep{2006ApJ...641...41L}.  We further take the shot noise to be dominated by quasars for which the effective number density is given by $\bar n \equiv [\int dL \phi(L) L]^2/ [\int dL \phi(L) L^2]$, where $\phi(L)$ is the full redshift evolution quasar luminosity function of \citet{hopkins07a}.\footnote{See \citet{meiksin18} for a discussion of observational uncertainties in $\bar n$, who finds that it is relatively well constrained at redshifts we consider.}  The effective number densities and source biases enter in the cross power spectrum of the ionizing source spatial overdensity between times $t_l$ and $t_m$ (see \citealt{meiksin18}):
\begin{equation}
P_S(k |  t_l,t_m) = c_{j, \delta}(t_l) c_{j, \delta}(t_m) P_\delta(k, t_l, t_m)+ \bar{n}^{-1} {\cal L}(t_l - t_m),
\end{equation}
where $P_\delta(k |  t_l,t_m)$ is the linear matter power at times $t_l$ and $t_m$, and ${\cal L}$ is the convolution of the source light curve with itself normalized so that ${\cal L}(0)=1$.  For simplicity, we assume quasars sources with top hat light curves with widths of $10, ~100,$ and $\infty~$Myr, with the former two values reflecting the range of estimates based on direct and indirect probes \citep{martini01}.

For the mean opacity coefficient, we use $\bar \kappa_\nu = 0.027 [(1+z)/5]^{5.4} (\nu/\nu_{1\rm Ry})^{-1.5}$ physical~Mpc$^{-1}$, for which the inverse of $\bar \kappa_{\nu = 1\rm Ry}$ (which yields the mean free path at the Lyman-limit) uses the measurement of \citet{worseck16} to $z=2.3-5.5$, and we have assumed the expected scaling with $\nu$ for an \HI\ column density distribution with slope of $\beta = 1.5$.  At lower redshifts than $z\approx 2$, opacity effects become unimportant as the propagation of radiation backgrounds becomes limited by the Horizon allowing us to use this fit even beyond its intended redshift range.  We take the response of the opacity to the ionizing background to be $c_{\kappa, \Gamma} =0.5$.  This value is consistent with an analytic models based on \citet{miralda00} for $\beta = 1.5$ and with what is directly measured from simulations at $z\sim 2-4$ \citep{mcquinn11}.\footnote{The parameter $c_{\kappa, \Gamma}$ enhances the amplitude of fluctuations by $(1-c_{\kappa, \Gamma})^{-1}$ on scales larger than the mean free path in the limit that all photons are absorbed in a time $<<H^{-1}$ (applicable at $z\lesssim 2$), leading to at most a factor of $2$ enhancement for the cases we consider.  We note that \citet{pontzen14a} used a value of $c_{\kappa, \Gamma} =0.9$ that is inapplicable to the IGM at considered redshifts, which results in a more significant enhancement.}

The remaining bias coefficients $ c_{\kappa, \delta}$ and  $c_{j, \Gamma}$, are set to zero. Setting $ c_{\kappa, \delta}$  to zero is justified because the sources are more biased than the absorbers.  The limit  $c_{j, \Gamma}= 0$ is applicable if the photoionization rate does not have a significant effect on modulating the emission: models find associated recombinations contribute at the $10-20$\% level to $\bar I_\nu(z)$ \citep{faucher09}.

To solve for $\delta_\Gamma$, we can average Equation~\ref{eqn:linInu} over frequency with weighting $w_\nu = \sigma_\nu$, and angle, using the solution to the homogeneous equation for $\bar I_\nu$.  

Solving this simpler frequency-averaged equation is found to be very accurate \citep{meiksin18}. We indicate the $\sigma_\nu \bar I_\nu$-weighted average of coefficients with a subscript $\sigma$.   Noting that the angular-averaged Green's function for the right hand side of the resulting equation is $j_0(k \eta_{t, t'}) \exp[-\int_{t'}^{t} dt'' \bar  \kappa_{\rm eff, \sigma}(t'')]$, where $\eta_{t, t'} \equiv \int^t_{t'} dt''/a''$ and $\bar  \kappa_{\rm eff, \sigma} \equiv (2 + \alpha_{\rm bk}) H + \Gamma^{-1} 4\pi \int d\nu \sigma_\nu \bar I_\nu \bar \kappa_\nu $, and ignoring fluctuations in the spectral index of the background on $\delta_\Gamma$, the solution to the spatial Fourier transform of this equation can be written in terms of the following transcendental equation
\begin{eqnarray}
  \label{eqn:deltaG1}
 \tilde \delta_{\Gamma}(t) &= \int_0^t dt' j_0 \left(k \eta_{t, t'}\right)\exp[-\int_{t'}^{t} dt'' \bar \kappa_{\rm eff, \sigma}(t'')] \\
& \times \left[    \tilde \beta_{0,\sigma} + \beta_{\delta, \sigma} \tilde \delta + \beta_{J, \sigma}  \tilde \delta_\gamma \right]. \nonumber
\end{eqnarray}
This equation can be solved for $\delta_{\Gamma}$ by discretizing the first integral in Equation~\ref{eqn:deltaG1} and some matrix algebra \citep{meiksin18}.  

\begin{figure*}
\begin{center}
\epsfig{file=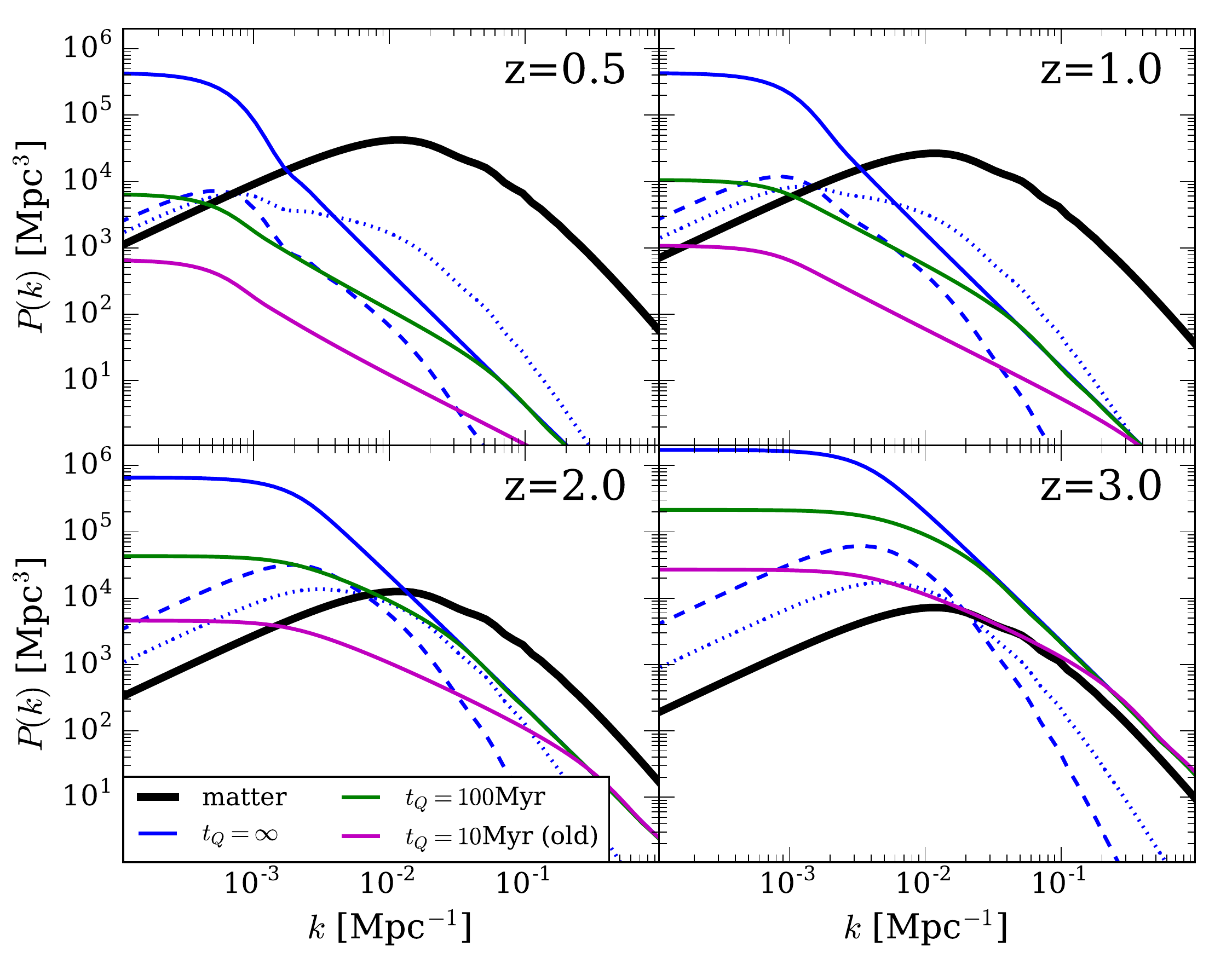, width=13cm}
\end{center}
\caption{Models for the power spectrum of the photoionization rate overdensity, $P_\Gamma$, at the specified redshifts. The solid colored curves show the shot noise contribution to $P_\Gamma$ for different quasar lifetime models.  The dashed blue curve is the clustered component of $P_\Gamma$, which does not depend on the quasar lifetime.  For reference, the solid black curve is the linear matter overdensity power spectrum.  The dotted curve is the cross power between the density tracing (clustered) component of the $J$ fluctuations and the linear matter overdensity; we find the dominant contribution from $J$ fluctuations to galaxy clustering owes to this cross correlation. }
\label{fig:Jfluc}
\end{figure*}
 
 Figure~\ref{fig:Jfluc} shows the results of the power spectrum of photoionization rate overdensity, $\delta_\Gamma$, fluctuations at $z=0.5,~1,~2,$ and $3$ for the different lifetime models.  The solid black curves are the matter density power spectrum, whose value (to order-unity density bias factors) indicates the level of fluctuations in LSS surveys.  The dashed blue curves are the density-tracing component of the model (which do not differ between lifetime models). The dotted curve is the cross power between the density-tracing (clustered) component of the photionization rate fluctuations and the matter density, the ionizing background term that often makes the largest imprint on galaxy clustering. The amplitude of fluctuations in the radiation background increases with redshift.  Even at low redshifts, the fluctuations can be larger than the density fluctuations at the lowest, horizon-scale wavenumbers that are of much interest for primordial non-Gaussianity searches.\footnote{The effect of radiation background for modes with $k \lesssim H^{-1}$ should be calculated in full General Relativity for (gauge independent) observables (as has been done for galaxy surveys; \citealt{2009PhRvD..80h3514Y}), which has not been done here.  However, we do not expect predictions for the size of effects to be altered (see \citealt{pontzen14a}).}  The importance of ionizing background fluctuations for LSS surveys depends on the bias with which these surveys trace the background fluctuations, $b_J$.  The ensuing sections estimate $b_J$ for several surveys.

\section{21cm intensity mapping surveys} 
\label{sec:HI}
Many efforts are coming online that aim to detect the post-reionization redshifted 21cm emission from the residual \HI\ gas that is trapped in and around galaxies.  These efforts generally do not have sufficient collecting area to detect individual sources, but instead will map the diffuse intensity of many sources (a mode referred to as `line intensity mapping').  Namely, the CHIME and HIRAX experiments are targeting $21$cm emission from $z=0.8-2.5$ \citep{2014SPIE.9145E..22B, 2016SPIE.9906E..5XN}, and MEERKAT is targeting this signal from $z=0.4-1.4$.  BINGO and FAST could extend the range of redshifts probed to $z=0$ \citep{2012arXiv1209.1041B, 2016ASPC..502...41B}, and mapping all the way to reionization is possible, perhaps with the Square Kilometer Array \citep{ska,2017arXiv170909066K} or with the instruments envisioned in \citet{cosmicvisions}.

Fluctuations in the ionizing background will modulate the distribution of \HI: Regions exposed to a larger ionizing background self-shield at higher densities and, hence, retain less \HI. Cosmological simulations post-processed with ionizing radiative transfer have been found to reproduce the broad features of the observed \HI\ column density distribution \citep{altay11, mcquinn11, rahmati13}.  Such simulations can be used to estimate the response of \HI\ to the background by post-processing simulations with different backgrounds and observing the change in \HI. We can in-effect do such a calculation by using the physically-motivated fitting formulae in \citet{rahmati13} that describe the density and photoionization rate dependences of the \HI\ fraction in their radiative transfer simulations.  
 Using these formulae, we find a factor of $\delta_{\Gamma}$ fractional change in the background results in a surprisingly large factor of $-0.25\delta_{\Gamma}$ change in the global amount of neutral hydrogen for $\delta_{\Gamma}\ll1$.  This estimated response does not depend on redshift to good approximation (being invariant to the extent that the column density distribution is invariant).  The details of the \citet{rahmati13} formula and this calculation are presented in Appendix~A. This large response translates to the intensity bias of $b_J = -0.25$ (see Eqn.~\ref{eqn:Pg}).

In addition to deriving $b_J$ from the relations in \citet{rahmati13}, we have pursued two other methods for estimating $b_J$ for 21cm intensity mapping surveys that result in similar values. Each of these methods make different assumptions. In one such method, we perform radiative transfer on a slab of a given \HI\ column with width corresponding to the Jeans scale at a specified density for $10^4$~K gas (see Appendix~\ref{sec:slab}). This Jeans-scale ansatz is motivated by the arguments in \citet{schaye00} and by the densities of absorbers in cosmological simulations \citep{altay11,mcquinn-LL}, and this ansatz results in a one-to-one relation between \HI\ column density and density. The motivation for this slab calculation is that it more explicitly tracks the ionization physics that is hidden in the \citet{rahmati13} fitting formulae (at the expense of the simplified geometry).  We find that a factor of $\delta_{\Gamma}$ fractional change in the ionizing background results in a factor of $-0.20\,\delta_{\Gamma}$ change in the global amount of neutral hydrogen, consistent with our previous estimate. Finally, one can do a simple estimate assuming that all absorbers have the same power-law profile and self shield at a critical density (\citealt{miralda00, mcquinn-LL}; see Appendix~\ref{sec:HI1}).  This model has been found to have some success at reproducing the shape of the observed column density distribution \citep{2002ApJ...568L..71Z}.  For a power-law profile that results in an \HI\ column density distribution with a power-law index of $\beta =1.3 (1.5)$ that is consistent with observations, this cruder model results in an even somewhat larger response of $-0.45$ ($-0.75$).  While we trust this model the least, the assumptions it makes are much different than the other model, and it still results in a large response.  The estimated value of $b_J$ is also redshift independent in both the slab and power-law absorber models.  All three models of $b_J$ ignore the effect on local radiation on the \HI\ ionization state, which could act to decrease the sensitivity to the ultraviolet background.  Simulations that include local radiation suggest that it is a minor factor in setting the ionization state \citep{rahmati13b}.

\citet{wyithe09} estimated a considerably smaller response of ($b_J\sim10^{-2}-10^{-3}$) of mean \HI\ density to the ionizing background.   Their estimate approximated DLAs as an exponential disks and ignored optically thin gas, leading to a smaller response than our estimates.  \citet{wyithe09} commented that $b_J$ would be of the order of unity if the density profile of self shielding systems was isothermal (i.e. $n_{\rm H} \propto r^{-2}$). Indeed, the \HI\ column density distribution suggests an underlying profile closer to isothermal \citep{miralda00, 2002ApJ...568L..71Z,mcquinn-LL}, and an isothermal-like profile was the basis of our third estimation method.

In addition to the intensity bias, $b_J$, to compute the amplitude of fluctuations also requires an estimate of the \HI\ density bias, $b_g$.  At $z=0.8$, the \citet{2013MNRAS.434L..46S} 21cm-galaxy cross-correlation measurement of $\Omega_{\rm HI} b_{g}  = [0.62^{+0.23}_{-0.15}] \times 10^{-3} $ combined with the quasar absorption line measurements of $\Omega_{\rm HI} = 0.5\times10^{-3}$ yield $b_g = 1.0\pm 0.3$.  At $z=2.1$, DLA clustering measurements find $b_g= 2.0\pm 0.1$ \citep{2012JCAP...11..059F, 2018MNRAS.473.3019P}.\footnote{ We note that DLA bias measurements are number density-weighted, as opposed to 21cm intensity mapping measurements, which weight by neutral hydrogen density, and so the  biases probed by these techniques should differ slightly. }
 There are currently no measurements of $b_g$ at higher redshift; our calculations at $z=3$ assume $b_g=2$.
 
 \begin{figure}
\begin{center}
\epsfig{file=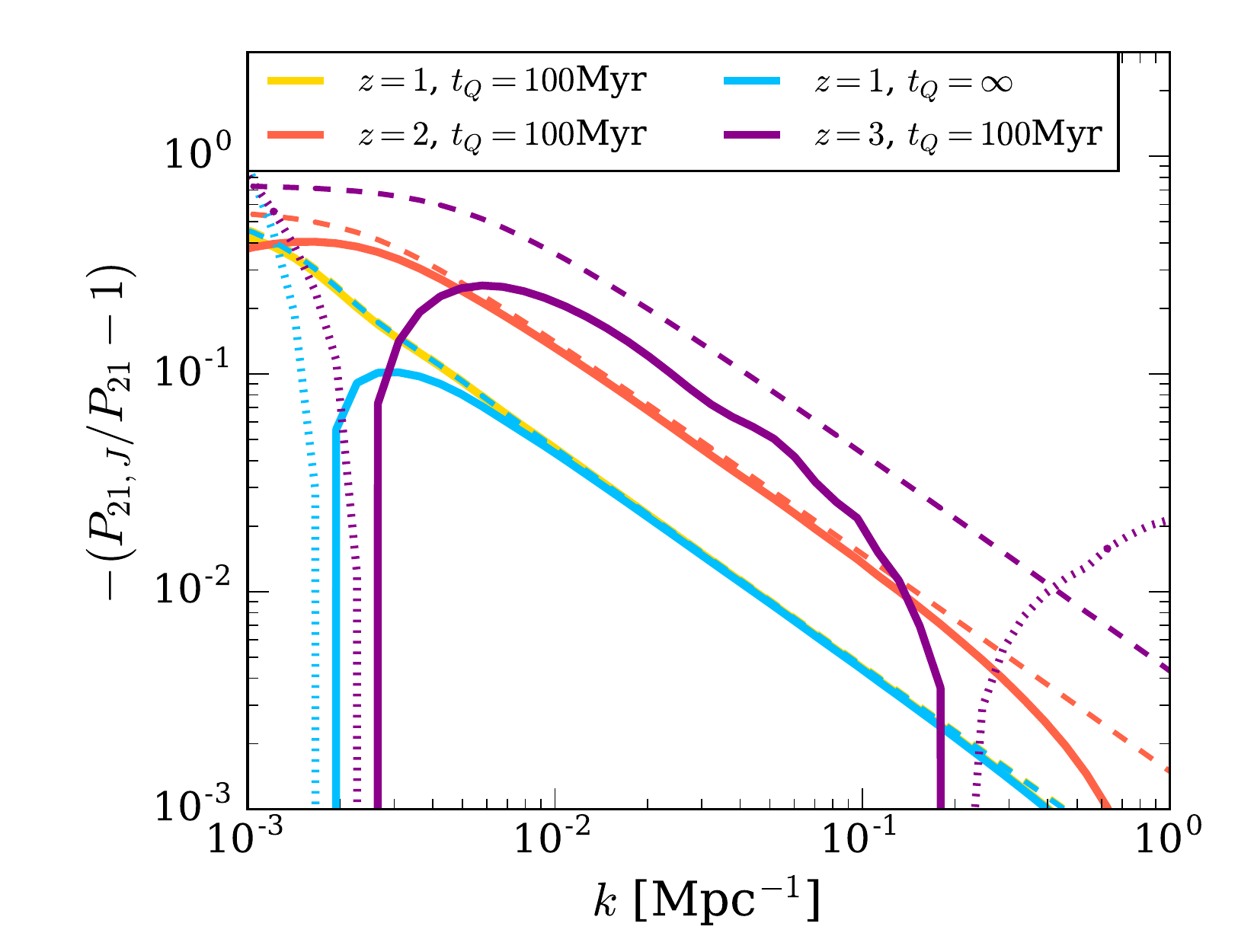, width=8cm}
\end{center}
 \caption{Predictions for the fractional impact of ionizing background fluctuations on the angularly-averaged power spectrum of the diffuse 21cm intensity, computed using $b_{J}=-0.25$, as motivated by the estimates discussed in the text. The solid curves are the negative of the total fractional contribution to the power from intensity fluctuations, and the dashed are the same but dropping the shot noise contribution to $\delta \Gamma$. The shot noise component becomes increasingly important with increasing redshift and increasing quasar lifetime. The dotted lines show the negative of the solid curves to show their behavior below zero.  \label{fig:21cm}}
 \end{figure}

Figure~\ref{fig:21cm} shows the fractional effect of intensity  fluctuations on the 21cm power spectrum at  $z=1$, $z=2$, and $z=3$ for models with quasar lifetimes of $t_Q = 100$ Myr and $t_Q = \infty$. For the intensity bias, we use our most detailed estimate based on \citet{rahmati13} that yields $b_J = -0.25$. The solid curves are the total contribution to the \HI\ power spectrum from intensity fluctuations, whereas the dashed curves are the contribution excluding shot noise (which depends on quasar lifetimes). The fractional effect of background fluctuations on the 21 cm power spectrum at $z = [1,2,3]$  is $\sim [0.03,0.1,0.3] (k/[10^{-2} \Mpc^{-1}])^{-1}$.

 The fractional effect of intensity fluctuations of the models that exclude shot noise scale as approximately $k^{-1}$, as the largest contribution from $J$ fluctuations owes to the $2 T_J P_\delta$ term with $T_J \sim k^{-1}$ at $k \gg \bar \kappa_{\rm eff, \sigma}$. This clustering component is comparable to the shot noise at the lowest and highest wavenumbers shown, most notably for the $z=3$ case. As quasars become more numerous toward $z\sim3$, the shot noise decreases, but the shot noise term is most prominent in our $z=3$ model because $b_J$ is larger as $z\rightarrow 3$ and this term goes as $b_J^2$.

We note that the lowest wavenumbers shown are difficult to measure by 21cm efforts due to foreground removal; foreground removal limits HIRAX to measurements to $k\gtrsim 0.03$~Mpc$^{-1}$ where the fractional imprint of intensity fluctuations is less than $10^{-3}, ~10^{-2}$ and $0.05$ at $z=1,~2$ and $3$ \citep{hirax}.  However, the projections for the \citet{cosmicvisions} are more optimistic, anticipating measurements to $k \approx 0.005 h$~Mpc$^{-1}$, a wavenumber where we predict fractional imprints of $\sim 0.1, ~0.3$ and $1$ at these redshifts (and the effects of intensity fluctuations would be even larger at the higher redshifts that \citet{cosmicvisions} is also targeting).

\section{recombination line surveys} 
\label{sec:rr}

Most recombination-line emissions originate from recombinations in the ISM of galaxies that result from the absorption of ionizing photons from nearby stars.  However, at $1 
\lesssim z\lesssim 3$ Type I AGN produce $\sim 10$\% of all ionizing photons, with stars producing the rest \citep[e.g.][]{haardt12}. Because most if not all of the ionizing photons from Type I AGN escape their host galaxies, but those of stellar origin likely do not, this suggests that $\sim$10\% of recombinations are tied to the ionizing background in the steady state limit in which recombinations are in global balance with emissions (which holds at $z\gtrsim 3$).  
As much of ten percent of photons of large-scale structure surveys that probe recombination lines, such as \HI\ H$\alpha$ and \HI\ Ly$\alpha$, could be from recombinations that trace the ionizing background.
  
 There are several upcoming Ly$\alpha$ and H$\alpha$ surveys whose cosmological determinations could be biased by these fluctuation-tracing recombinations. HETDEX aims to constrain the Cosmology from the clustering of Ly$\alpha$ emitters at $1.9 <z<3.5$ \citep{2008ASPC..399..115H}, and SPHEREx -- a proposed NASA Medium-class explorer satellite that is funded for concept studies -- aims to detect H$\alpha$ emission from galaxies from much of the Cosmic Volume \citep{2016arXiv160607039D}.  HETDEX and SPHEREx will undertake both `traditional' galaxy surveys and intensity mapping campaigns (with the latter defined as surveys that map all the emissions and do not locate individual sources). The effect of ionizing backgrounds is largest for intensity mapping surveys as we find below that traditional galaxy surveys likely do not detect the majority of the (diffuse) background-sourced recombination photons.

 To estimate the size of the background-tracing emissions in these lines, let us decompose the recombination radiation emission into an internal-to-galaxies component -- driven by ionizing photons absorbed before they escape their host ISM and the subsequent recombination -- and an external component -- driven by ionizing photons in the extragalactic ionizing background:
\begin{eqnarray}
\label{eq:jlya}
j_{\rm rec} &=&  j_{\rm rec, int} + j_{\rm rec, ext},  \\
           & = &   \bar j_{\rm rec, int} (1 + \delta_{g}) + \bar j_{\rm rec, ext}(1+ \delta_{g} + c_{\rm rec,\Gamma} \delta_\Gamma) \nonumber,
 \end{eqnarray}
 In the first line, $j_{\rm rec, int}$ and $j_{\rm rec, ext}$ are the internal emission coefficient and external (background-sourced) emission coefficient for the desired recombination line. The second line expands the two emission sources into an overdensity that traces the sample galaxies, $\delta_g \equiv (b_g + f \mu^2) \delta$ and, for the external coefficient, also an overdensity that traces the ionizing background $\delta_\Gamma$ with bias $c_{\rm rec,\Gamma}$. To the extent that the cross sections of absorbers are unchanged by the ionizing background and that every photoionization is balanced with a recombination -- approximations that are likely to hold -- $c_{\rm rec,\Gamma}=1$.

From Equation~\ref{eq:jlya}, we can compute the galaxy power spectrum for a luminosity-weighted galaxy clustering measurement or an intensity mapping survey:
\begin{eqnarray}
P_{\rm rec}(k) &=&  
\left[ (b_{g}+f\mu^2) 
+ \overbrace{f_{\rm ext} c_{\rm rec,\Gamma}}^{b_J} T_J \right]^2 P_{\delta_L}(k) \nonumber \\  
&+& f_{\rm ext}^2 c_{\rm rec,\Gamma}^2 n_J(k, \nu)^{-1}.
\end{eqnarray}
where, to connect to our previous notation (c.f.~Eqn.~\ref{eqn:Pg}), $b_J = f_{\rm ext} c_{\rm rec,\Gamma}$ and we have defined $f_{\rm ext} \equiv  \bar j_{\rm rec, ext}/(\bar j_{\rm rec, int} + \bar j_{\rm rec, ext}) \approx  \bar j_{\rm rec, ext}/\bar j_{\rm rec, int}$.  We note that at $z\gtrsim 3$, when photons that make it into the background are quickly absorbed and, thus, sourcing recombinations, $f_{\rm ext}$ is given by the fraction of ionizing photons that escape galaxies -- including ones hosting AGN -- ($f_{\rm esc}$) times the fraction of these ionizing photons that will be absorbed ($\approx \kappa_\sigma/\bar{\kappa}_{\sigma,{\rm eff}}$)  times the fraction of these recombinations the survey picks up ($f_{\kappa,g}$):
$$f_{\rm ext} \approx \frac{\bar j_{\rm rec, ext}}{\bar j_{\rm rec, int}} \approx  f_{\kappa,g} \times \frac{\bar{\kappa}_\sigma}{\bar{\kappa}_{\sigma,{\rm eff}}} \times f_{\rm esc}. $$

Our calculations use the quasar emissivity function of \citep{khaire15} and the UV-only star formation rate from \citet{haardt12} to calculate $f_{\rm ext}$, making the assumption that all ionizing photons escape from quasars, and approximately none make it out of galaxies.  We find $f_{\rm esc} = [0.06, 0.07, 0.04]$ and $\bar{\kappa}_{\sigma}/\bar{\kappa}_{\sigma,{\rm eff}} = [0.12,0.42,0.70]$ at $z=[1,2,3]$, values we adopt for subsequent calculations. 

\begin{figure}
\begin{center}
\epsfig{file=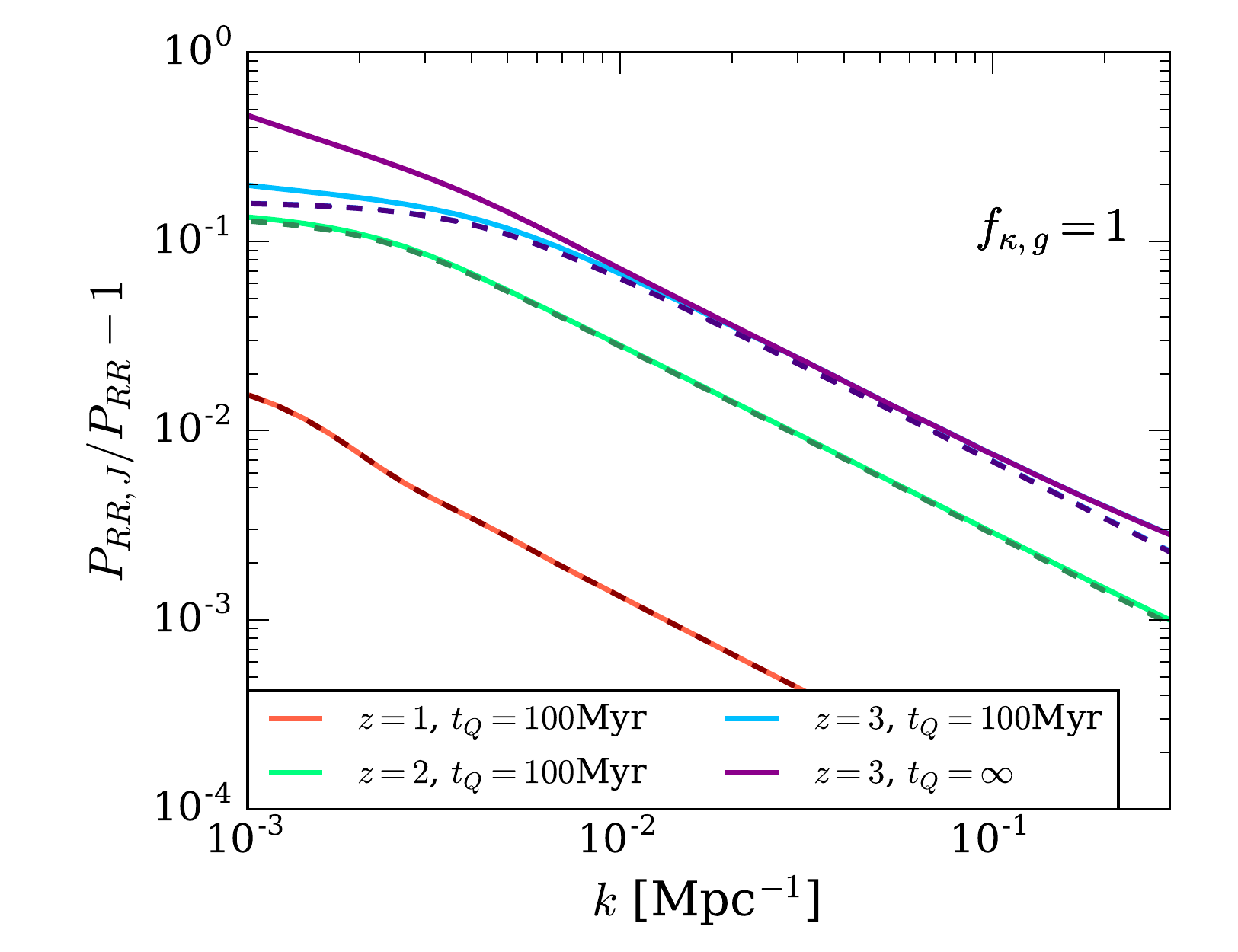, width=8cm}
\end{center}
 \caption{Predictions for the impact of ionizing background fluctuations on intensity mapping surveys targeting recombination lines (such as \HI\ Ly$\alpha$ and H$\alpha$).  The solid curves are the total fractional contribution to the angularly-averaged power spectrum from intensity fluctuations, and the dashed is the contribution excluding shot noise.  While these estimates are for the intensity mapping mode, the main text also considers recombination surveys that selects individual sources (where the effect of backgrounds is smaller).   \label{fig:recomb} }
\end{figure}

\subsection{Intensity mapping surveys}

Figure~\ref{fig:recomb} shows our estimates for the recombination-line flux power spectrum at $z=1$, $z=2$, and $z=3$ for ionizing background fluctuation models with quasar lifetimes of $t_Q = 100$ Myr and $t_Q = \infty$.  These estimates are for the intensity mapping mode in which $f_{\kappa,g} = 1$ so that all recombination photons (emitted towards the observer) contribute to the survey.  This applies, for example, to the intensity mapping surveys of SPHEREx in H$\alpha$ and HETDEX in Ly$\alpha$.  For these calculations, we use the estimated Ly$\alpha$ intensity density bias of $b_g = [1.16,1.19,1.40]$ for $z = [1,2,3]$ of \citet{2014ApJ...786..111P}, and if we assume the same $b_g$ values should apply for H$\alpha$ in intensity mapping then our calculations also hold for this line (an assumption which should hold up to preferential dust destruction of Ly$\alpha$ photons in more massive galaxies).  We further assume that the sizes of absorbers are not affected by $\Gamma$ so that $b_{\Gamma} = 1$, and, since $b_J \equiv f_{\rm ext} b_{\Gamma}$ and using the previously quoted $f_{\rm ext}$, we find  $b_J = [0.007, 0.03, 0.03]$ at $z=[1,2,3]$. The uncertainty in our estimate of $b_J$ comes primarily from $f_{\rm esc}$: Though measurements of photoionization rates in the Ly$\alpha$ forest constrain the total number of ionizing photons, the number of ionizing photons per unit star formation is still unknown to a factor of $\sim2$ \citep{shull12}, as is the star formation rate density and the fraction of recombination photons that are not absorbed by dust, leading to an uncertainty in our estimates of the order of unity.

 The solid curves are the total fractional effect to the power spectrum from intensity fluctuations, and the dashed curves show the fractional effect without shot noise. Note that the intensity bias for recombination radiation is positive, in contrast to our 21cm intensity bias, so the effect of intensity fluctuations is to enhance the power.  The clustering component of intensity fluctuations (and really its cross with the survey galaxy overdensity) is much more important than the shot, even somewhat more so than for 21cm surveys because our estimated $b_J$ is smaller for recombination surveys.  We estimate that the fractional contribution to the power of intensity fluctuations scales as $k^{-1}$ with normalization ranging from $0.005$ to $0.1$ at $k=10^{-2}$Mpc$^{-1}$.

\subsection{Traditional galaxy surveys}

The fluctuations are smaller for a traditional galaxy survey compared to ones in intensity mapping mode that we just discussed, as only a fraction of the recombinations associated with background ionizations occur in the aperture used to measure the light from surveyed galaxies so that $f_{\kappa,g}\ll 1$ is likely rather $f_{\kappa,g}= 1$ for intensity mapping. An estimate for $f_{\kappa,g}$ is
\begin{equation}
f_{\kappa,g} \approx \frac{A_{\rm ins}n_{g}}{\bar \kappa_{ \sigma}},
\end{equation}
where $A_{\rm ins}$ is the aperture of the instrument and $n_{g}$ is the number density of surveyed galaxies. For HETDEX, using $\sigma_{\rm ins}= 1.8$~arcsec$^2$, we find a very small value of $f_{\kappa,g}\approx 1\times10^{-4}$ for $z=2$, using $\bar{\kappa}_{ \sigma}^{-1}=240$ pMpc and a Ly$\alpha$ emitter density of $n_{g} = 9\times10^{-5}$ Mpc$^{-3}$. For the SPHEREx galaxy survey, with $\sigma_{\rm ins}= 38$~arcsec$^2$, for $z=1,2$, using $\bar{\kappa}_{ \sigma}^{-1}= 650,240$ pMpc, and ${\bar n_g} = 1.5\times10^{-3}$ Mpc$^{-3}$, we estimate $f_{\kappa,g}\approx 0.1,0.06$. Because $f_{\kappa,g}$ is so small, for the galaxy-selecting campaigns of HETDEX and SPHEREx the component of the clustering that traces the background will be smaller than our intensity mapping estimates for these surveys by $f_{\kappa,g}$, making the effect of intensity fluctuations largely irrelevant. The imprint from ionizing fluctuations that could be more relevant to these campaigns is discussed in the following section.

\section{Other galaxy surveys}
\label{sec:galsur}

The ionizing background fluctuations may affect the properties of galaxies in general -- and not just their \HI\ fractions and the rate of recombinations, as considered in previous sections. Galaxy properties may be modulated by ultraviolet backgrounds, as the $ 13.6-200$eV background ionizes the gas, affecting the rate at which it can cool, and the $\approx11$eV Lyman-Werner background dissociates molecular hydrogen. The former effect has been observed in cosmological simulations \citep{efstathiou92,benson02,hambrick09,hambrick11}, though it is most significant in $\lesssim10^{10}\Msun$ halos. While for the larger galactic halos catalogued in large-scale structure (LSS) surveys the modulating effects from background fluctuations is likely small, a rough estimate for their magnitude is important as we target increasingly subtle cosmological imprints. 
 
 \subsection{Galaxies selected by star formation rate}
 \label{sec:sfr}
 
Most properties by which galaxies are selected for LSS surveys are related to the galactic star formation rate or the galactic stellar mass.  To estimate the effect of the ionizing background on the star formation rate, we start with the ansatz that the observability of a galaxy is proportional to the cooling time of its halo gas. This ansatz is motivated by the idea that star formation is tied to cooling and condensation in the halo \citep[e.g.][]{sharma12, voit17, mcquinn17, tumlinson17}. This approach fails to capture the nonlinearity of the physics governing star formation, but serves as an illustrative first-order estimate. 

Processes such as stellar feedback, metal enrichment, and changes in gas temperature from an increased background intensity make the relation between background intensity and star formation potentially very complex. However, in the picture where galaxies are relatively closed systems with minimal energy escape (which is a reasonable assumption at low redshift) and each galactic system is an approximate steady state, any additional cooling would be balanced by additional star formation its associated feedback to reheat the gas and balance cooling.  A more detailed picture might be found by running simulations of individual galaxies exposed to different ionizing backgrounds and measuring the resulting change in brightness. This approach would come with its own set of caveats -- the CGM in such simulations is often unable to reproduce observed properties, such as commonly observed ions, with the simulated CGM depending sensitively on how feedback is prescribed. 

With our simple model, we investigate the cases of galaxies with virial temperatures of $5\times 10^{5}$, $\sim 10^6$ and $5\times10^{6}$K.  Such temperatures correspond to halos of approximately $10^{11.5}-10^{13}$M$_{\odot}$ at $z\sim0$, for which abundance matching techniques find host stellar masses of $\sim 10^{10}-10^{11}$ M$_{\odot}$ (\citealt{behroozi10,li09}) -- typical star-forming galaxy masses. To calculate cooling rates, we assume that gas in these halos is likely to have densities near the halo `virial density' or mean halo gas density ($\bar{\rho}_{\rm halo} = 200\bar{\rho}$, where $\bar{\rho}$ is the cosmic mean gas density).  Models suggest typical densities of halo gas at $\sim 100~$kpc from the galaxy that range from roughly the mean halo density to a factor of ten smaller \citep{sharma12, mcquinn17, fielding17}.  In addition, the gas that actually cools and condenses, sourcing star formation, might be the denser gas that lies closer to galaxies.  Thus, we discuss a range of densities centered around the virial density.

To understand how the ionizing background can shape halo cooling, we have run a grid of {\sc Cloudy} ionization models for virialized halo gas.  The cooling rate of gas depends on its temperature $T$, spectral shape, and ionization parameter, $U$, where $U=\Phi/n_{\rm H} c$. In these estimates, $n_{\rm H}$ is the density of hydrogen nuclei, and $\Phi$ is the ionizing photon flux, which is proportional to the photoionization rate, $\Gamma$. We use the rates from the \citet{haardt12} model; the metagalactic ionizing background should dominate the ambient ionization radiation experienced by halo gas \citep{mcquinn17, uptonsanderbeck18}. 
Because we are concerned with a fractional change in cooling and because metals dominate the cooling rate at these temperatures for both collisionally and photoionized gas, variations in metallicity do not change the results. 
Our fiducial ionizing background model assumes a photon flux, $\Phi= 4.8\times10^{4}$ cm$^{-2}$ s$^{-1}$, at $z=0.2$ based on \citet{haardt01}, and a spectral shape based on \citet{haardt12}. Our calculations also assume that the spectral shape is the fluctuations is also unchanged at all relevant frequencies, which should largely hold at $z<2$ when the ionizing background is limited by  light travel.

\begin{figure}
\begin{center}
\epsfig{file=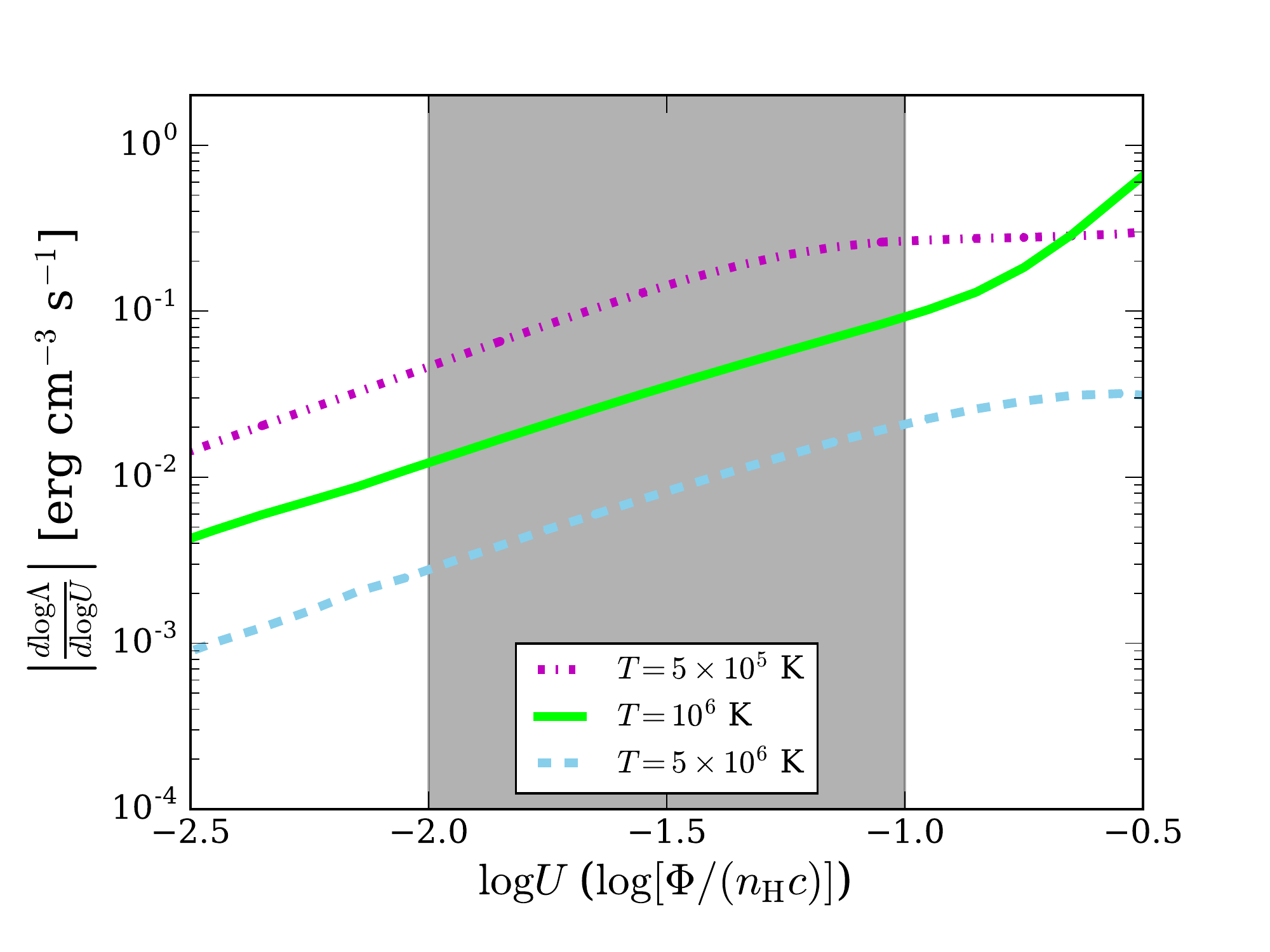, width=9cm}
\end{center}
 \caption{The fractional change in the cooling rate with respect to a fractional change in the ionization parameter, which is equal to our intensity bias parameter $b_J$. The purple, green, and blue curves show this change for gas temperatures of $5\times 10^{5}$, $10^6$ and $5\times10^{6}$K, respectively. The grey band brackets the ionization parameters, corresponding to a factor of three less and more than the mean halo gas density of $z=0.5-2$ halos ($\log{U}\approx-1.5$ for the mean halo gas density at all of these redshifts). \label{fig:cool}}
\end{figure}

Figure~\ref{fig:cool} shows how the cooling rate changes as a function of ionization parameter based on these {\sc Cloudy} computations. The curves represent the specified gas temperatures. The shaded band represents the range of ionization parameters if the halo gas is within a factor of three of the mean halo density $\bar{\rho}_{\rm halo}$. This same shaded band applies over all considered redshifts $z=0.5-2$, as the mean halo density happens to change almost inversely with photoionization rate at these redshifts such that $\Phi/n_H$ is surprisingly constant. The fractional change in the cooling rate of the virial temperature curves in this allowed band range between $\sim 10^{-3}$ and $10^{-1}$. (The fractional change is much less than one because the ionization state of the hot halo gas is set largely by collisions and, hence, less shaped by the ionizing background.)

As in the two previous sections, Equation~\ref{eqn:Pg} describes the power spectrum of galaxies where the crucial piece we aim to calculate is the intensity bias, $b_J$.  Under our assumption that the observability of a galaxy is proportional to the cooling time of its halo gas, then $b_J = d\log \Gamma /d\log U$, the quantity calculated in Figure~\ref{fig:cool}. While the exact $b_J$ in this model is dependent on gas temperature and strongly on ionization parameter, let us consider a fiducial model with gas temperature of $10^6$ K and an ionization parameter representing gas just at the virialized density at $z=0.5-2$, such that $\log{U} = -1.5$. This results in $b_J=-0.05$. This estimate for $b_J$ is a factor of several larger or smaller if we vary the virial temperature or density by factors of a few, and our results for the fractional impact on the power spectrum depend linearly on $b_J$. 

The galaxy density bias, $b_g$ is calculated using a Sheth-Tormen halo mass function \citep{sheth01}, where $b_g$ is $[1.24,1.47,2.12]$ for $z = [0.5,1.0,2.0]$ at a virial temperature of $10^6$ K corresponding to halo masses of $[1.5\times 10^{12},1.1\times 10^{12},6.2 \times 10^{11}]$ M$_{\odot}$.\footnote{For $5\times10^5$ K gas, $b_g=[1.06,1.23,1.72]$ at $z = [0.5,1.0,2.0]$, corresponding to halo masses of $[5.2\times 10^{11},3.6\times 10^{11},2.1 \times 10^{11}]$ M$_{\odot}$.}

\begin{figure}
\begin{center}
\epsfig{file=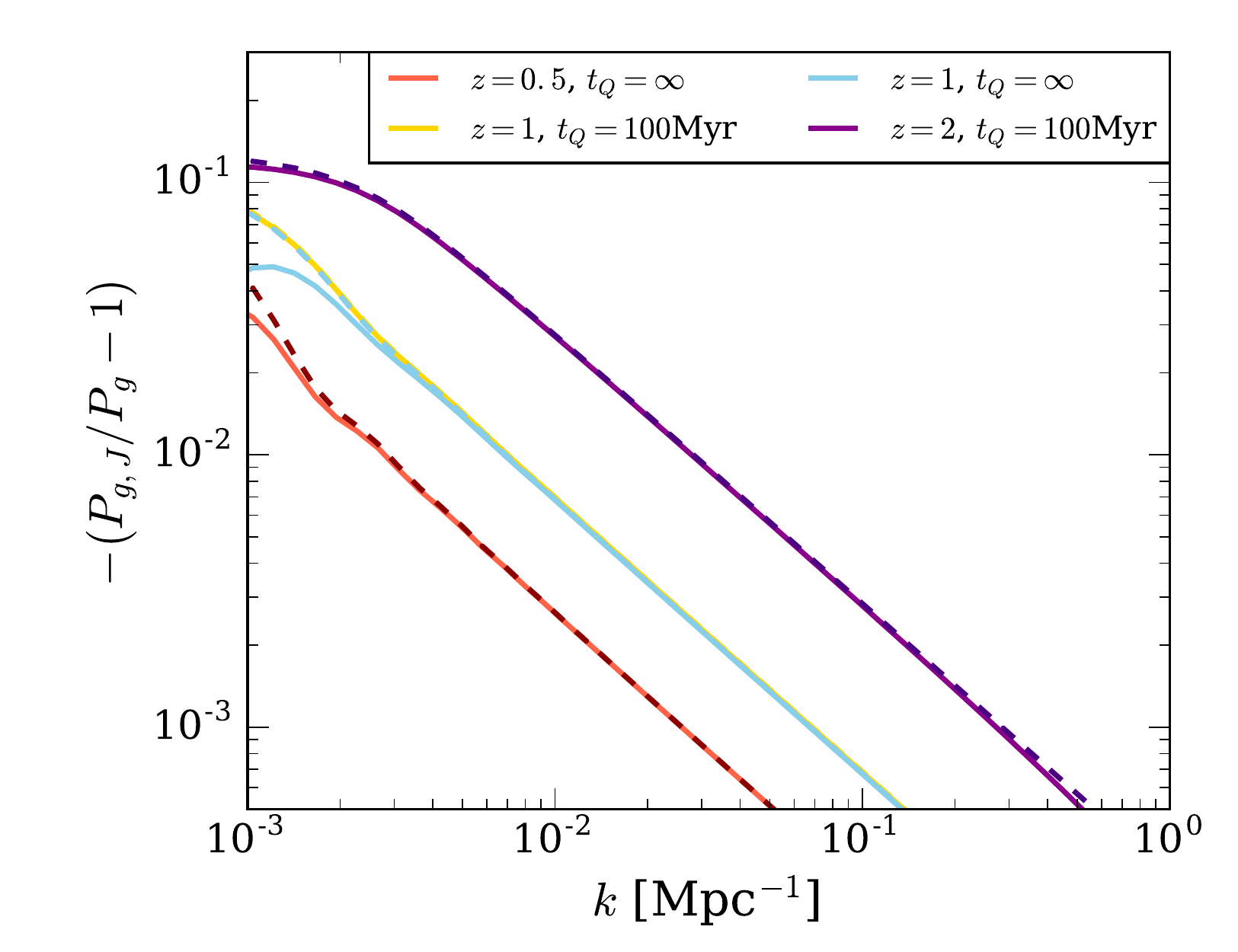, width=9cm}
\end{center}
\caption{Predictions for the fractional change from ionizing background fluctuations on the $z=0.5$, $z=1$ and $z=2$ galaxy power spectrum for $b_J= 0.05$.  Unlike in the previous plots, these estimates apply to galaxy surveys of `typical' galaxies with $T_{\rm vir}\sim10^6$K and not just to intensity mapping surveys.  However, the values are more uncertain as these calculations require assumptions for how galaxy formation is affected by ionizing backgrounds. Our model that connects observability to halo cooling suggests $b_J=[2\times 10^{-3}-0.2]$; the amplitude of the fractional change in the power spectrum scales linearly with $b_J$. The solid curves are the total contribution to the power spectrum from intensity fluctuations, and the dashed curves with corresponding colors are the contribution excluding shot noise.   \label{fig:gc}}
\end{figure}

Figure~\ref{fig:gc} shows the effect of fluctuations on the galaxy flux power spectrum under these choices for $b_g$ and $b_J$. Because the intensity bias is small, the effect of fluctuations is no more than a couple percent at $k<10^{-2}$ Mpc$^{-1}$. However, precision measurements in future surveys (such as searches for primordial non-Gaussianity at the sub-percent level) may still be biased by this small modulation.  We estimate by how much this level of contamination could bias cosmology constraints in the following section. 
 
A galaxy's star formation rate could also be modulated by metagalactic Lyman-Werner ($\sim 11$eV) background, which can dissociate molecular hydrogen and hence prevent star formation.  The effective mean free path of such backgrounds is smaller than this for ionizing backgrounds ($\sim 0.1 H^{-1}$ compared to $\sim H^{-1}$ at low redshift) so that fluctuations in the Lyman-Werner background are larger. However, the Lyman-Werner background inside a molecular cloud is likely to be dominated by galactic star formation at kpc-scale separation \citep{2010ApJ...721L..79G}, unlike the background that sets the ionization of the much more extended circumgalactic medium at 100 kpc.  Thus, the modulation of total galactic star formation rates from Lyman-Werner backgrounds is likely to be smaller than from ionizing backgrounds, which can affect gas accretion onto a galaxy. 

 \subsection{Galaxies selected by stellar mass}
 \label{sec:stellarmass}

 The previous subsection's estimates were for galaxy selection that is sensitive to star formation rate. If the selection is more sensitive to stellar mass, the background that shapes the cooling is more weighted towards the past, when the fluctuations were larger, predicting a larger imprint of the background.  Additionally, in the past, background fluctuations could have a larger effect on cooling because of the lower virial temperatures of the progenitor system. One could also imagine that if the background were larger in a region, then there would be less cooling at earlier times leading to larger densities and more cooling at later times, reducing the response we have estimated.

We note that clustering measurements based on number density weighting of galaxies will be affected by the number of galaxies around the detection threshold. If the magnitude of the ionizing background were increased, a galaxy that may otherwise have been above this threshold could be undetectable due to a slight modulation in luminosity and conversely, an otherwise fainter galaxy may be slightly more luminous in the presence of a diminished background. The magnitude of this effect depends on the slope of the cumulative luminosity function at the detection threshold as this determines the number density of galaxies that would come into or out of the survey. The effect of number density weighting results in $b_{J} \rightarrow 2.5 \alpha b_{J, \rm th}$, where $b_{J, \rm th}$ is our previously calculated $b_J$ but evaluated at the survey detection threshold, $\alpha \equiv d\log_{10} n(>m_{\rm th})/dm$ where $ n(>m)$ is the cumulative distribution above some apparent magnitude $m$ with $m_{\rm th}$ being the threshold. Values of $\alpha$ are typically quoted at $\alpha \sim0.5-2$ \citep{liu14,menard10}, although likely the enhancement is modest for a deep survey like SPHEREx. For our case where $b_{J, \rm th} = 0.05$, number weighting would result in somewhat larger values of $b_J \approx 0.06-0.25$ if $\alpha = 0.5-2$.

\section{Cosmological parameter biases}
\label{sec:nongaus}

\begin{figure}
\begin{center}
\epsfig{file=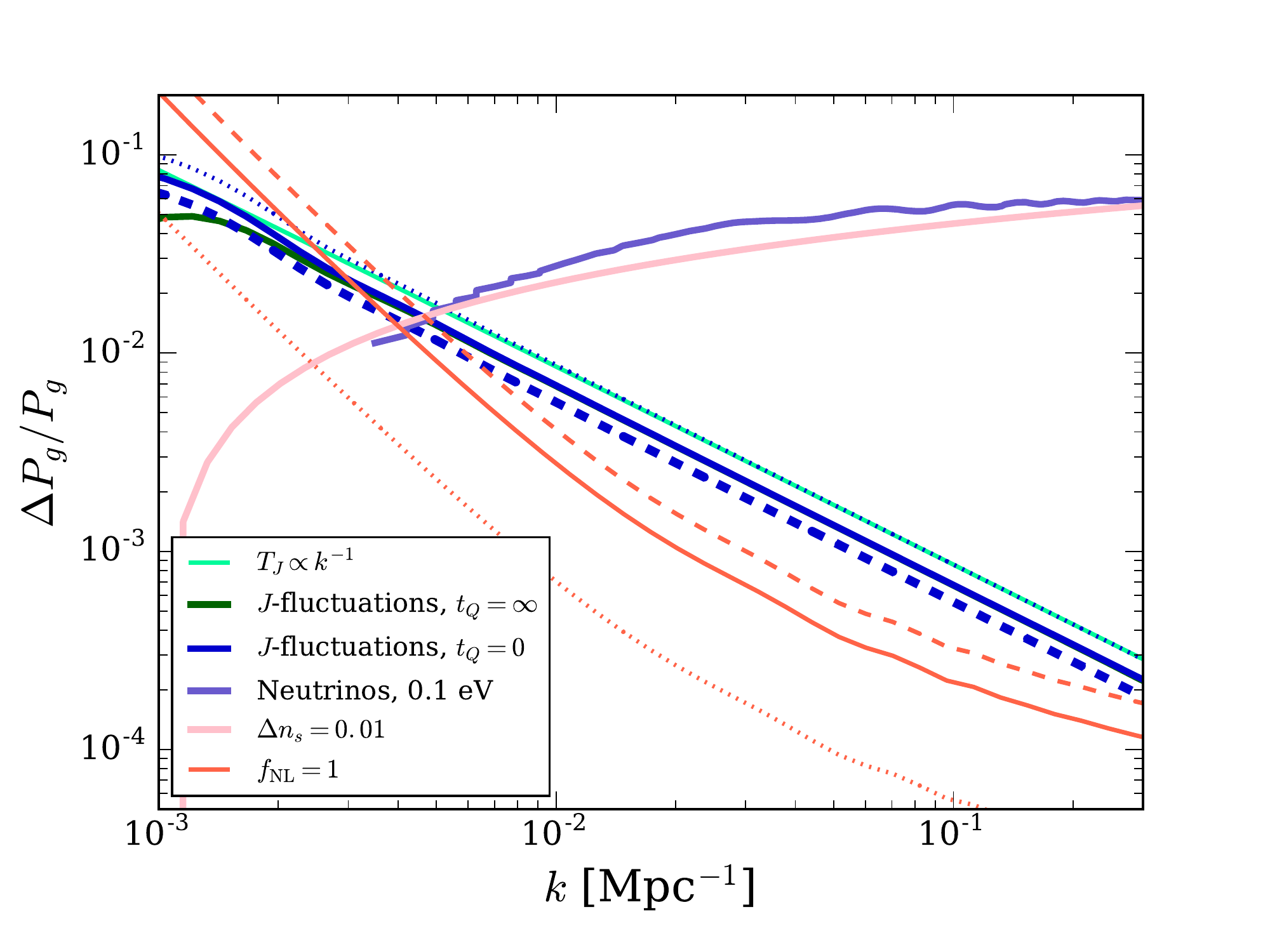, width=9cm}
\end{center}
 \caption{Residuals of different cosmological effects on the galaxy power spectrum at $z=1$. The dark green curve shows the absolute value of the fractional change in the power spectrum of our model from Section~\ref{sec:galsur} with background fluctuations sourced by quasars with infinite lifetimes. The dark blue curves of different line styles show the same model, excluding shot noise with $b_g = [1.1,1.5,1.9]$ (dotted, solid, dashed). The thin turquoise curve shows our background fluctuations model excluding shot noise with a modified transfer function that scales as $k^{-1}$. The fractional effect on power from neutrinos with a total mass over all flavor eigenstates of 0.1 eV is shown in purple, $\Delta n_s= 0.01$ is shown as the pink band, and $f_{\rm NL}=1$ is shown in orange with $b_g = [1.1,1.5,1.9]$ (dotted, solid, dashed). \label{fig:deltap}}
 \end{figure}

Finally, we investigate the robustness with which certain cosmological parameters can be constrained from galaxy redshift surveys in the face of the contamination from ionizing background fluctuations. Figure~\ref{fig:deltap} shows the $z=1$ galaxy fractional effect on the power spectrum of the fluctuation models from the previous section.  The dark green curve shows the full model with $t_{Q}=\infty$ and the dark blue dashed curves show just the clustering component only. For the latter, we show models with three different galaxy biases, $b_g = 1.1,1.5,1.9$, with this range motivated in the calculations below.  We want to compare these to motivated benchmarks for some of the most interesting cosmological parameters. In particular, also shown is the fractional effect of neutrino mass of $0.1$ eV on the $z=1$ galaxy power spectrum relative to the massless case (computed with CAMB; \citealt{Lewis:2002ah}), of a change in the spectral index of scalar potential fluctuations with $n_s \pm 0.01$, and of the squeezed triangle non-Gaussianity parameter with $f_{\rm NL}=1$ for a comb of bias models again with $b_g = 1.1,1.5,1.9$ relative to the gas with $f_{\rm NL}=0$. We use the expression in \citet{dalal08} for the scale dependent bias (see also \citealt{matarrese08,slosar08}).  The values of these parameters are motivated by being comparable or better than current CMB limits \citep{planck18} and being at the level achievable with future galaxy redshift surveys \citep{baldauf16}.

The shape of the residual fractional imprint from massive neutrinos and $n_s$ variations are different than background fluctuations, and, therefore, we expect the effect of background fluctuations are unlikely to be confused with these effects, though could still introduce bias if not modeled. However, the shape induced by a finite $f_{\rm NL}$ is qualitatively similar to background fluctuations.  A survey that aims to estimate the value of $f_{\rm NL}$ may be biased if background fluctuations are not included and marginalized over. 

To estimate this bias (and the cost of such marginalization) we use a Fisher Matrix approach \citep{tegmark97}, which provides a quick way to estimate the inverse covariance matrix (the Fisher Matrix) of parameter constraints from a measurement (in our case a measurement of $P_g$).  This formalism assumes that second-order terms of the Taylor expanded log likelihood (${\cal L}$) of the survey power are sufficient to estimate the errors.  The Fisher Matrix of the parameters $p_i$ for a galaxy survey measurement of the power spectrum $P_g(k,z)$ is 
\begin{equation}
    F_{ij} \equiv -\Big\langle \frac{\partial^2 {\cal L}}{\partial p_i \partial p_j} \Big\rangle = \sum_k \frac{1}{\sigma_P^2} \frac{\partial P_g}{\partial p_i} \frac{\partial P_g}{\partial p_j},
    \label{eq:Fij}
\end{equation}
where the sum in wavenumber goes over all power spectrum wavenumber bins (also known as ``band powers''). 
  The error on a band power owes to a combination of cosmic variance and shot noise:
\begin{equation}
    \sigma_P^2 = \frac{4\pi^2}{k^3 \Delta(\log{k}) V(z)} \left(P_g(k, z) + \frac{1}{{\bar n_g}(z)} \right)^2,
    \label{eq:sigmaP}
\end{equation}
where $V(z)$ is the survey volume, $\Delta(\log{k})$ is the width of the band power bin, and ${\bar n}_g$ is the galaxy number density. We evaluate the comoving volume for a given redshift as $4\pi \chi(z)^3/3$, where $\chi(z)$ is the comoving distance to redshift $z$. This enables us to explore how the estimated values of $f_{\rm NL}$ change for a survey that would observe the entire comoving volume at a given redshift, but performing the parameter estimation on $k> k_{\rm min}$, where the minimum wavenumber depends on the survey's geometry and also can be shaped by survey systematics (like diffuse light from the galaxy).

We want to estimate the bias on $f_{\rm NL}$ for a survey that ignores the effect of background fluctuations.  The Fisher matrix allows us to estimate the bias on each parameter $p_i$ via   
\begin{equation}
    \Delta p_i 
    = \sum _j F^{-1}_{ij} \left( \sum_k \frac{1}{\sigma_P^2}\frac{\partial P_g}{\partial p_j} \Delta P_g \right),
    \label{eq:dpi}
\end{equation}
where $\Delta P_g$ is the unaccounted effect of intensity fluctuations on the measured galaxy clustering power spectrum.  This expression for the bias follows from the optimal quadratic estimator \citep{1998PhRvD..57.2117B,1998ApJ...503..492S,2000ApJ...533...19B}.

In the simplified analysis presented here to understand the bias for measuring $f_{\rm NL}$, we consider a two-parameter model where $P_g$ is parameterized by the galaxy density bias, $b_g$, at each redshift and by $f_{\rm NL}$ (i.e. the set of the $p_i$ is comprised of only these two parameters). We fix all of the standard cosmological parameters to fiducial values -- given by \citep{2016A&A...594A..13P} -- as we do not expect that including these in the calculation would change the estimated bias since these parameters have a smaller effect on low wavenumbers affected by intensity fluctuations and $f_{\rm NL}$. 

\begin{figure}
\begin{center}
\epsfig{file=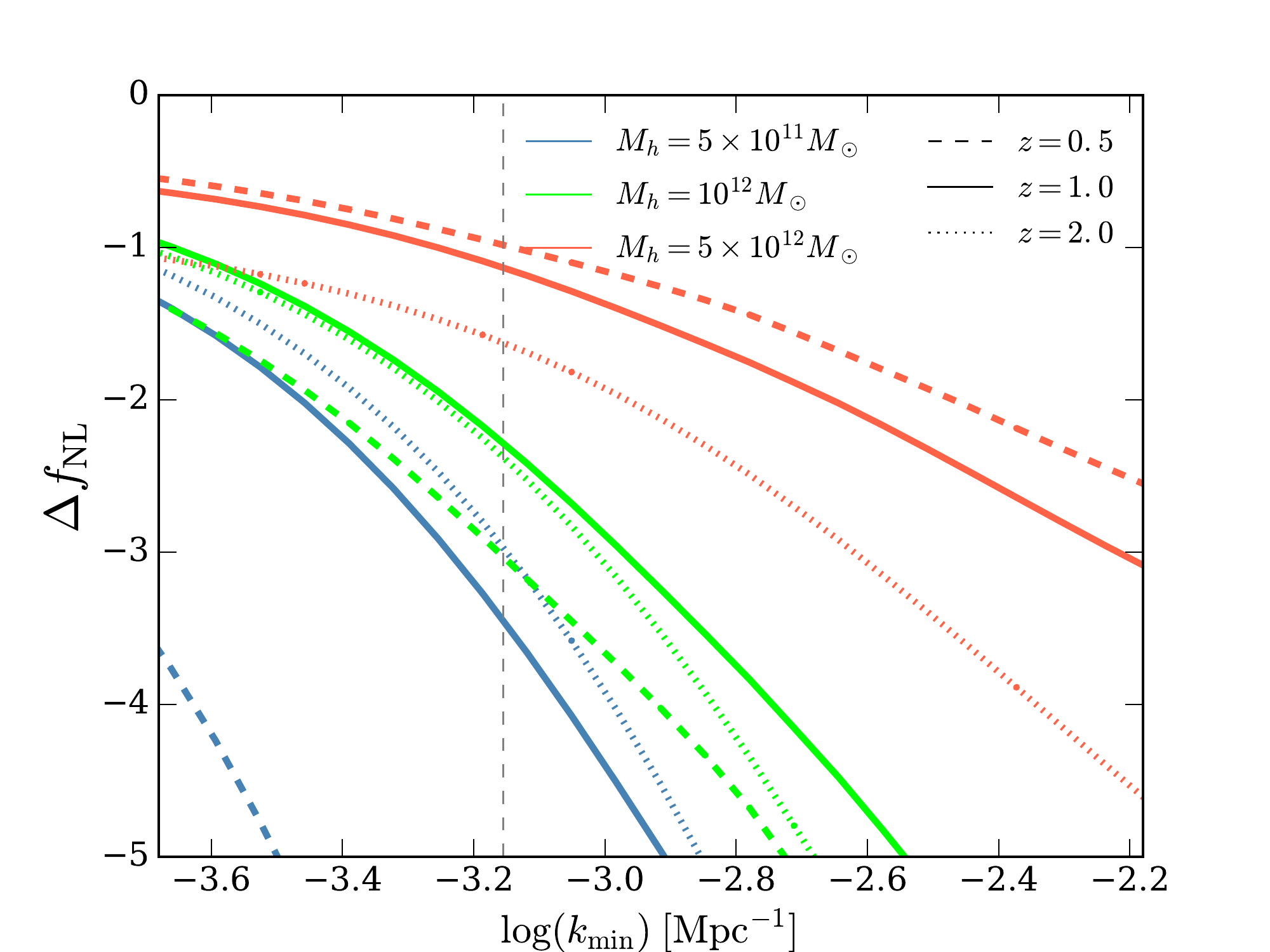, width=9cm}
\end{center}
 \caption{Bias in the estimate of $f_{\rm NL}$ from a survey which neglects intensity fluctuations as a function of the largest scale probed by the survey, $k_{\rm min}$. The dashed, solid and dotted curves are $z=0.5,~1.0$, and $2.0$, respectively.  Different colours correspond to the values for this bias, $\Delta f_{\rm NL}$, when assuming different halo masses (and thus different galaxy biases and number densities -- see text for details). The vertical dashed line shows the typical $k_{\rm min} = 10^{-3}\;h~\mathrm{Mpc}^{-1}$ that is assumed in the forecasts of \citet{SPHEREX}. The bias of intensity fluctuations used is $b_J = -0.05$. The resulting $\Delta f_{\rm NL}$ depends linearly on the assumed $b_J$. \label{fig:biasFNL} }
 \end{figure}

Figure~\ref{fig:biasFNL} shows how the bias in $f_{\rm NL}$ parameter, $\Delta f_{\rm NL}$, depends on the largest mode probed by a given survey ($k_{\rm min}$). A value of $b_J = -0.05$ was used for the calculation, but this value can be smaller if the response of the cooling rate to photoionizing background is suppressed. However, in the more likely case of number density weighting or if the halo gas is colder or lower density we found it could be larger. Different colors show the value of $\Delta f_{\rm NL}$ for different values of the halo mass: $5\,\times 10^{11}\,M_\odot$ (blue), $10^{12}\,M_\odot$ (green) and $5\,\times 10^{12}\,M_\odot$ (red). Using a Sheth-Tormen halo mass function \citep{sheth01}, we can match halo mass to both the number density of galaxies and galaxy bias. Moreover, such a formalism predicts the redshift evolution of the galaxy bias and number density. For redshifts [$0.5, 1.0, 2.0$], as considered in this paper, the biases and number densities are: $b_g(z) = [1.05, 1.29, 2.02]$ and ${\bar n_g}(z) = [3.9, 3.7, 2.6]\,\times\,10^{-3} \;(h/\mathrm{Mpc})^3$ for the halo mass of $5\,\times 10^{11}\,M_\odot$; $b_g(z) = [1.16, 1.45, 2.33]$ and ${\bar n_g}(z) = [2.1, 1.8, 1.1]\,\times\,10^{-3} \;(h/\mathrm{Mpc})^3$ for the halo mass of $10^{12}\,M_\odot$; and, $b_g(z) = [1.52, 2.00, 3.39]$ and ${\bar n_g}(z) = [0.41, 0.31, 0.11]\,\times\,10^{-3} \;(h/\mathrm{Mpc})^3$ for the halo mass of $5\,\times 10^{12}\,M_\odot$. These estimates depend on the smallest wavenumber used to constrain $f_{\rm NL}$, with its bias $\Delta f_{\rm NL}$ increasing with increasing $k_{\rm max}$. In Figure~\ref{fig:biasFNL}, and unless otherwise stated, we have assumed $k_{\rm max} = 0.2\;h/\mathrm{Mpc}$.

Here we discuss the trends and dependencies in the $f_{\rm NL}$ bias, $\Delta f_{\rm NL}$. The smaller the survey -- and hence the larger the value of $k_{\rm min}$ -- the larger $\Delta f_{\rm NL}$, a trend which owes to the relative effect of intensity fluctuations becoming larger towards smaller scales compared to the signal from $f_{\rm NL}$. Moreover, the smaller the halo mass, the larger the effect on $\Delta f_{\rm NL}$. This trend arises from the larger bias of more massive halos and because the effect of $f_{\rm NL}$ is via an additive contribution to the galaxy bias that scales as $ f_{\rm NL} (b_g - 1)$.  Since bias enhances the signal from non-Gaussianity with respect to intensity fluctuations, the net result is a linear bias that $ \Delta f_{\rm NL} \propto \, (b_g - 1)^{-1}$ and that depends on wavenumber. When the density bias approaches unity, the non-Gaussianity signal vanishes, and $\Delta f_{\rm NL}$ approaches infinity. (In the same regime, however, the estimated error on the $f_{\rm NL}$ parameter ($\sigma_{f_{\rm NL}}$) increases towards infinity. Since both signals depend on this scale-dependent bias contribution to $b_g$, their dependence on $(b_g-1)$ largely cancels out, and the ratio of $\Delta f_{\rm NL}/\sigma_{f_{\rm NL}}$ remains finite across the transition of $b_g=1$.)  The value of $\Delta f_{\rm NL}$ in our calculations also mildly depends on the number density of galaxies, ${\bar n_g}$. However, as long as the number density is high enough that the shot-noise contribution in Equation~\ref{eq:sigmaP} does not dominate on large scales, then the value of $\Delta f_{\rm NL}$ does not vary with ${\bar n_g}$. 
Finally, as long as the amplitude of intensity fluctuations is small, i.e. $b_J \ll 1$, then $\Delta f_{\rm NL} \propto b_J$ and so the bias increases linearly with $b_J$.

This bias can be mitigated by marginalizing over possible intensity fluctuation models. When marginalizing over the amplitude $b_J$ assuming the shape of the intensity fluctuation imprint is known, we find that the errors on $f_{\rm NL}$ increase by [53, 42 and 26]\% at $z=[0.5,1,2]$ respectively. 
 While the shape of $T_J$ is mostly determined by the clustered component of intensity fluctuations, whose wavenumber scaling is known on scales smaller than the mean free path.  However, the shape is somewhat affected by the stochastic nature of the sources. To test how the exact shape of $T_J$ influences the results, we have replaced the shape of $T_J$ with a simple $k^{-1}$ scaling for the residuals that is matched visually to our fiducial model at a wavenumber of $k_p = 0.002\;h\mathrm{~Mpc^{-1}}$ (see Figure~\ref{fig:deltap}). (In this model, the transfer function $T_J$ becomes $T_J(k_p) (k/k_p)^{-1}$.) Using this simple approximation for $T_J$, we recomputed $\Delta f_{\rm NL}$.  We find that this approximation reduces the bias $\Delta f_{\rm NL}$ considerably, by a factor of $0.1$, $0.2$ and $0.3$ at redshifts of 0.5, 1.0 and 2.0 respectively.  In conclusion, while ignoring the effect of UV fluctuations can potentially bias the estimated value of the $f_{\rm NL}$, including the suggested simple template and marginalizing over $b_J$, considerably reduces bias at the cost of a $\sim 40\%$ larger error on $f_{\rm NL}$.

\section{Conclusions}
Motivated by the principle that the ionizing background is the only non-gravitational astrophysical process that can affect $\gg1$ Mpc large-scale structure (LSS) correlations, this paper made estimates for the impact of ultraviolet background fluctuations for various galaxy surveys.  In particular, we have focused on efforts that map the diffuse post-reionization 21cm intensity, that select millions of  Ly$\alpha$ and H$\alpha$ emitters, and that survey Milky Way-like galaxies. 

For the diffuse 21cm radiation from all galaxies, a larger ionizing background results in less \HI.  While most cosmic \HI\ resides in damped Lyman-$\alpha$ systems that are self shielded to the ionizing background (and, hence, seemingly well protected from fluctuations), we developed three different estimates that suggest that the response of the \HI\ fraction is not insignificant, with our two most realistic estimates finding a $0.25 \delta \Gamma/\Gamma$ change, where $\delta \Gamma$ is the infinitesimal change in the  background photoionization rate. We further find that this response is relatively constant with redshift. We also developed a model for fluctuations in the ionizing background, which, in combination with the response estimates, were used to predict the effect of background fluctuations on galaxy clustering. We estimated a fractional change in the 21cm power spectrum of $\sim [0.03,0.1,0.3] (k/[10^{-2} \Mpc^{-1}])^{-1}$ at $z = [1,2,3]$ relative to the case without UV background fluctuations. 

We next investigated surveys that target recombination lines, such as Ly$\alpha$ and H$\alpha$ emitter surveys. A larger ionizing background results in an increase in recombination emission, as the number of recombinations is tied to the number of photoionizations.  When surveying in intensity mapping mode (to capture recombinations that may occur far from a galaxy), the intensity bias is roughly the ratio of ionizing photons that make it into the background to those that are produced in total. For $z = [1,3]$, we find a fractional response of $[0.008,0.03] \times \delta \Gamma/\Gamma$.  With our model for fluctuations in the ionizing background, we estimated a fractional change in power of $\sim [0.001,0.1] (k/[10^{-2} \Mpc^{-1}])^{-1}$ at $z = [1,3]$ relative to the case without UV background fluctuations. However, for the campaigns that correlate individual galaxy positions (rather than intensity mapping the diffuse background), we find the imprint of intensity fluctuations is likely to be negligibly small. 

Finally, we considered how a standard galaxy survey could be affected by ionizing backgrounds.  Following work that connects the cooling rate in galactic halos to their star formation rate, we adopted the ansatz that the star formation rate is proportional to the cooling rate of halo gas.  Considering motivated ranges for the density of the relevant cooling gas in halos (centered around $200$ times the mean density), for gas at $10^6$ K we found a fractional response in the star formation rate of $\sim 0.01-0.1\delta \Gamma/\Gamma$, and a resulting fractional change in the galaxy clustering power spectrum of $\sim [10^{-3},10^{-2}] (k/[10^{-2} \Mpc^{-1}])^{-1}$ for luminosity weighting.  Number weighting can increase the effects by a factor of a few as well as somewhat smaller temperatures, and the fractional impact increases with redshift owing to the ionizing background fluctuations being larger.  Because these calculations relied on the simple ansatz that galaxy observability scales with the cooling of halo gas (and the properties of diffuse halo gas are also poorly constrained), this estimate is more uncertain than our estimates for 21cm and recombination radiation surveys.

For typical responses, the modulation from ionizing backgrounds is generally dominated by the cross correlation between the traditional galaxy clustering signal and the density-tracing component of intensity fluctuations, resulting in a fractional imprint that typical scales as $k^{-1}$ on wavenumbers larger than the inverse of the photon mean free path. We also considered the stochastic contribution to intensity fluctuations from the rareness of the sources.  We found that the stochastic term can matter if (1) $b_J\gtrsim 0.1$ resulting in a large coupling to intensity fluctuations and (2) quasars with $> 100$ Myr lifetimes are the dominant contributor to the ionizing background. The former condition was only satisfied for our estimates for 21cm intensity mapping surveys.

The effect of ionizing backgrounds is the only astrophysical contaminant of cosmological correlations on greater than megaparsec scales (as winds from galaxies only travel a limited distance).\footnote{Galactic feedback can affect low wavenumbers in the galaxy power spectrum from the large-scale manifestation of small-scale behavior.  Such effects are captured by the effective coefficients in the effective perturbation theory of large-scale structure \citep{2015JCAP...05..019L} and should not inhibit large-scale analyses.}  We compared the spectrum and amplitude of the predicted effects to benchmark cosmological parameter constraints targeted by forthcoming large-scale structure surveys, finding that varying neutrino mass or the scalar spectral index, $n_s$, effect the galaxy clustering power spectrum with a much different spectral imprint than intensity fluctuations.  However, the effects of intensity fluctuations are more similar to local primordial non-Gaussianity. We find that measurements of the squeezed-triangle primordial non-Gaussianity parameter  $f_{\rm NL}$ using the galaxy power spectrum could be biased by fluctuations in the ionizing background at the level of $\Delta f_{\rm NL}  \sim (1-3) \times (b_J/0.05)$, near benchmark values for surveys such as SPHEREx. However, the value of $|b_J|$ can be smaller than $0.05$ if the response of the cooling rates in galaxies to the intensity fluctuations is suppressed, or larger when considering the selection bias from the number density weighted surveys. This would reflect in smaller, or larger, value of $\Delta f_{\rm NL}$. Our calculations find that marginalizing over a template that scales as $k^{-1}  P_{\delta_L}(k)$, almost completely removes this bias, but comes with a cost of 40\% larger error bars on $f_{\rm NL}$.

In conclusion, as long as they are not ignored, intensity fluctuations are unlikely to  substantially hamper precision cosmology even with futuristic galaxy redshift surveys. 

\section*{Acknowledgments}
We would like to thank the anonymous referee for their thoughtful comments. We would like to thank Anson D'Aloisio, Andrew Pontzen, and Simeon Bird for detailed comments on an earlier draft. We also thank Sarah Tuttle for helpful discussions. This work was supported by United States NSF awards AST~1514734 and AST~1614439, by NASA ATP award NNX17AH68G, by NASA through the Space Telescope Science Institute award HST-AR-14575, and by
the Alfred P. Sloan foundation. PRUS was partially supported by NSF award number AST-1817256. AM acknowledges support from the UK Science and Technology Facilities Council.

\bibliographystyle{mnras}
\bibliography{References}
 
\appendix

\section{A. Models for the response of \HI\ to a change in the ionizing background}
In this appendix, we describe our three methods for computing the response of \HI\ to a change in the ionizing background as described in \S~\ref{sec:HI}.

\subsection{Model based on results of \citet{rahmati13}}
\label{sec:rahmati}

This method uses the \HI\ column density distribution from cosmological simulations post-processed with ionizing radiative transfer presented in \citet{rahmati13}. The calculations simulate the LSS, the properties of halo gas, feedback processes, as well as ionizing radiative transfer with complicated geometries.  \citet{rahmati13} calculated how the \HI\ in these simulations correlates with the incident ionizing background, finding that the photoionization rate as a function of density is given by the physically-motivated fitting formula
\begin{eqnarray}
\label{eq:gamma_H}
\frac{\Gamma_{n_{\rm H}}}{\Gamma_{\rm bk}} = 0.98 \left[1 + \left( \frac{n_{\rm H}}{n_{\rm H, SS}} \right)^{1.64} \right]^{-2.28} \nonumber\\ + 0.02  \left[1 + \left( \frac{n_{\rm H}}{n_{\rm H, SS}} \right) \right]^{-0.84},
\end{eqnarray}
where $\Gamma_{\rm bk}$ is the background photoionization rate incident upon the absorber, $\Gamma_{n_{\rm H}}$ is the photoionization rate which it self shields to, and $n_{\rm H, SS}$ is an estimate for 
the density of the IGM that starts to self-shield. This density threshold is given by setting $\tau_{\rm HI} = 1$ and using the following relation from \citet{schaye01}:
\begin{eqnarray}
\label{eq:NHInH}
N_{\rm HI}= 2.3\times 10^{13}{\rm cm}^{-2}\left(\frac{n_{\rm H}}{10^{-5}{\rm cm}^{-3}}\right)^{3/2}\left(\frac{T}{10^4 {\rm K}}\right)^{-0.26}\nonumber\\
\times\left(\frac{\Gamma}{10^{-12} {\rm s}^{-1}}\right)^{-1}\left(\frac{f_g}{0.16}\right)^{1/2},
\end{eqnarray}
which is derived by relating the size and density of absorbers by assuming they have size equal to the Jeans length.
Evaluating for $N_{\rm HI}=4\times10^{17}$, this gives
\begin{eqnarray}
n_{\rm H, SS} = 6.7\times10^{-3} \frac{\sigma_{\rm eff}}{2.49\times10^{-18} {\rm ~cm}^{-2}} \left(\frac{T}{10^4 {\rm K}}\right)^{0.17} \nonumber\\
\times\left(\frac{\Gamma}{10^{-12}{\rm s}^{-1}}\right)^{2/3} {\rm ~~cm}^{-3}.
\end{eqnarray}
These formulae allow one to relate the \HI\ fraction of a system to its density and the ionizing background intensity.

In addition to these formulae, our model requires as input the column density distribution in simulations in the case where the systems' \HI\ fraction is computed assuming the background is unattenuated, which we take from \citet{rahmati13}.  (This unattenuated column density distribution is more power-law like, not showing the `break' at $N_{\rm HI} \sim 10^{18}$cm$^{-2}$ that owes to self shielding.  We also use that this column density distribution does not change substantially with redshift \citep{prochaska09b}, though simulations suggest a mild steepening at $N_{\rm HI} \ll 10^{18}$cm$^{-2}$ and higher redshifts than considered here.

\begin{figure}
\begin{center}
\epsfig{file=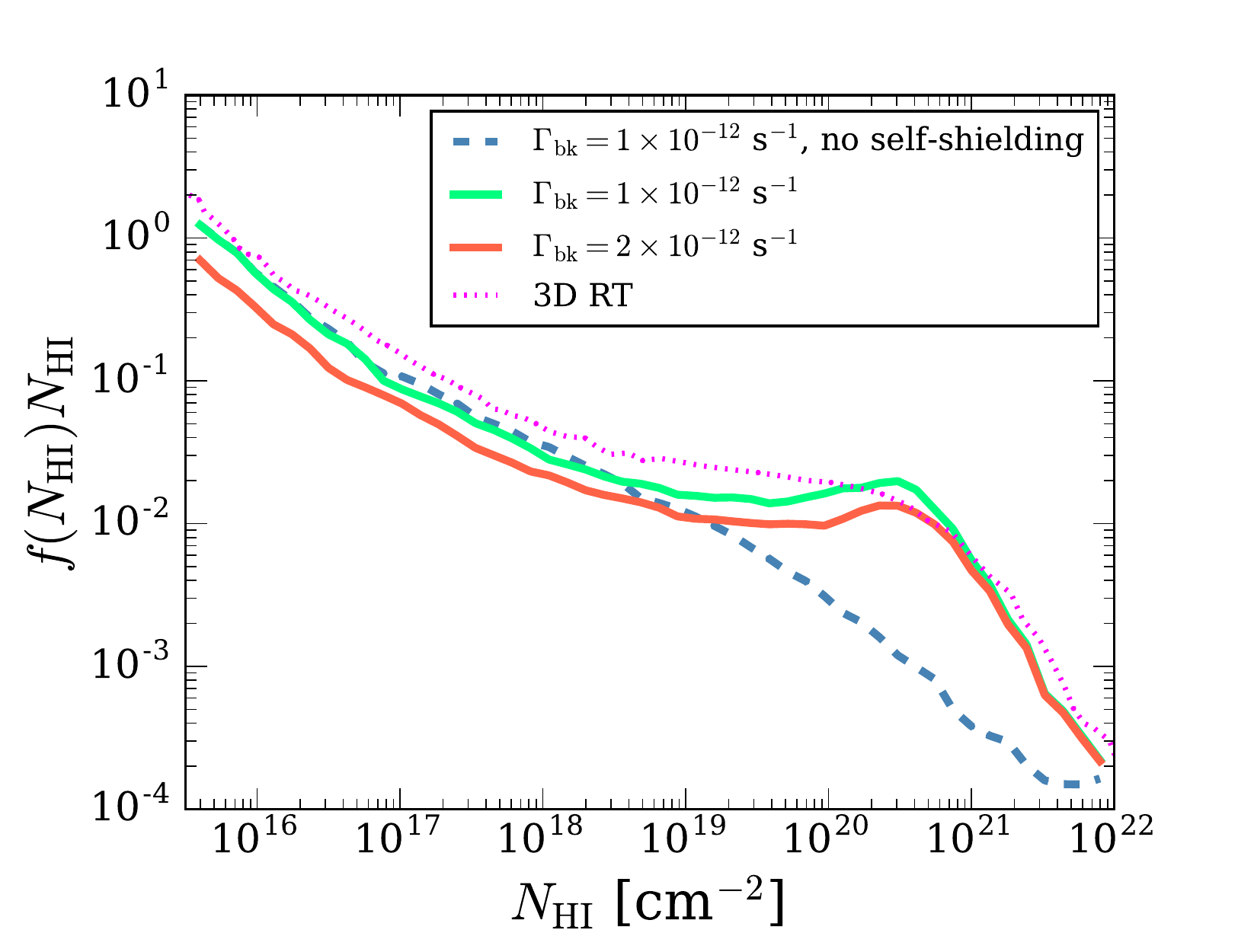, width=9cm}
\end{center}
 \caption{Column density distribution functions at $z=3$ calculated using our model and in \citet{rahmati13}. The blue curve assumes the background is everywhere unattenuated and uses cosmological simulations to compute \HI\ densities, and the magenta dotted curve is the same but computed using full 3D radiative transfer calculation with the same background to model the attenuation, from \citealt{rahmati13}.  The green and red solid curves are our calculations that include the effects of self-shielding from the \citet{rahmati13} fitting formulae presented in \S~\ref{sec:rahmati} for different photoionization rates. We use the calculations represented by the solid curves to estimate the response of \HI\ to an ionizing background.  \label{fig:fNhi}}
 \end{figure}

We utilize equation~(\ref{eq:gamma_H}) to calculate the photoionization rate, $\Gamma_{\rm HI}$, which then allows us to calculate the neutral fraction. Using the ratio of the shelf-shielded to the optically thin neutral fraction, we determine how changes in the background alter the self-shielding threshold and, subsequently, how changes in the photoionization rate affect the neutral fraction. Figure~\ref{fig:fNhi} shows the $z=3$  background-unattenuated column density distribution function from \citet{rahmati13} (dotted curve) as well as our calculation for the self-shielding column density distribution function using the \citet{rahmati13} fitting formulae (green curve). The magenta dotted curve is the full 3D self-sheilding calculation in a cosmological simulation, from \citet{rahmati13}.  Note that qualitatively the change is similar to our approximate calculation, suggesting that our estimate should be reasonably accurate.  The red curve shows how the column density distribution function changes when the global photoionization rate is doubled. 

Using the optically thin column density distribution in \citet{rahmati13}, a gas temperature of $10^4$ K, and ionizing background photoionization rate of \citet{}, we calculate the bias as the change in mean cosmic number density of \HI\ dictated by the magnitude of the ionizing background as a function of the photoionization rate and redshift,
\begin{equation}
b_J =  \frac{d\log }{d\log \Gamma} \left(\int dN_{\rm HI}^{\rm NSS} f^{\rm NSS}(N_{\rm HI}) N_{\rm HI}(N_{\rm HI}^{\rm NSS}| \Gamma,z) \right),
\end{equation}
where $f^{\rm NSS}$ is the column density distribution with no self-shielding. 

\subsection{Photoionized slab model}
\label{sec:slab}
Our second method to calculate how \HI\ responds to the ionizing background models absorbers as semi-infinite slabs.  This method lacks the complex 3D geometry of self-shielding systems that is encoded into the \citet{rahmati13} fitting formulae, but makes the ionization physics (which is enfolded in the \citet{rahmati13} fitting formulae) more transparent. These slabs are modeled as infinite in extent and their width is set by the Jeans length. We set the density $n_{\rm H}$ as given by Equation~\ref{eq:NHInH} for a given column, and we take the temperature to be $T=10^4$K.  We perform 1D radiative transfer calculations with the incident radiation oriented perpendicular to the slab, with one ray incident from both sides and the two rays summing to a total incident photoionization rate of $\Gamma_{\rm bk}$. The incident background spectrum is modeled as a simple power law of the form $J_{\nu}\propto (\nu/\nu_0)^{-\alpha}$, where $\alpha=1$ with a cutoff at $4~$Ry, motivated by the shape of ionizing background models above $1~$Ry \citep{haardt12}. The radiative transfer divides the slab into cells, attenuating the rays as they traverse cells by their neutral fraction during the previous iteration, casting one ray after the other. In each cell, we assume the gas within the cell is in photoionization equilibrium to calculate the neutral fraction that will be used for subsequent iterations. This calculation is then iterated until the neutral fraction has converged.

This calculation yields a mapping between our initial ``no self-shielding'' $N_{\rm HI}^{\rm NSS}$ and $N_{\rm HI}^{\rm SS}$, just as in our calculation based on \citet{rahmati13}.  Again using the optically thin column density distribution of \citet{rahmati13} we can transform to the optically thick using this calculation, and estimate the response of the \HI\ abundance. The final results of this slab model calculation yields a $0.20\delta_{\Gamma}$ change in the amount of \HI\ when the photoionization rate is increased by a factor of $\delta_{\Gamma}$ for $\delta_{\Gamma}\ll 1$.

This model and that in \S~\ref{sec:rahmati} assumed the size of the absorber is the Jeans length \citep{schaye00}.  The Jeans length is only an approximate scale and so, even to the extent this assumption holds, it should only hold to an order unity factor.  
  To understand the sensitivity of our models to this ansatz, we varried the slab width and find that the \HI\ bias changes negligibly ($\sim1$\%) when varying the assumed width within a factor of two around the Jeans scale.

\subsection{Model based on a critical density for self shielding and power-law profiles}
\label{sec:HI1}
Finally, a rough estimate for the impact of ionizing background fluctuations on the amount of \HI\ can be obtained from modeling the self-shielding absorbers as having a single power-law profile and a specific density $\Delta_{\rm SS}$ above which the gas self shields \citep{miralda00, furlanetto05, mcquinn-LL}.  Unlike the previous two approaches, this method does not rely on numerical simulations or the \citet{schaye01} model for the density of absorbers.  In addition, the approximation that self-shielding occurs at a critical density reasonably agrees with what is found in simulations \citep{mcquinn-LL}, and  this power-law model has some success at reproducing the column density distribution of Lyman-limit systems and DLAs \citep{2002ApJ...568L..71Z}.

In this model, the photoionization rate sets the value of $\Delta_{\rm SS}$ via the relation $\Gamma \propto \Delta_{\rm SS}^{(7 - \gamma)/3}$, where $\gamma$ is the power-law index of the gas density distribution $P(\Delta)$ at $\Delta \gg1$, where $\Delta$ is in units of the cosmic mean density \citep[see][]{mcquinn-LL}.  The mass in HI is proportional to $\int_{\Delta_{\rm SS}}^\infty d\Delta P(\Delta) \sim  \Delta_{\rm SS}^{1 - \gamma} \sim \Gamma^{\frac{3(1-\gamma)}{7-\gamma}} \sim \Gamma^{-3/2(1-\beta)}$, where $\beta$ is the spectral index of the column density distribution at optically thin columns ($\gamma$ and $\beta$ are easily related in this model).  Thus, the overdensity in \HI\ is related to the overdensity in $\Gamma$ via $\delta_{\rm HI}= -3/2(1-\beta) \delta_\Gamma$ meaning that $b_J = 3/2(1-\beta)$ in Equation~\ref{eqn:Pg}. For $\beta = \{1.3, 1.5, 1.7\}$, the range consistent with observations at $N_{\rm HI}\sim 10^{17}$cm$^{-2}$,  we find $\{-0.45, -0.75, -1.05\}$.  This somewhat larger response than the $-[0.2-0.3]$ yielded by our other methods, again suggests that the response is large.  The response is likely somewhat weaker than this model predicts because this model does not capture the self-shielding transition well, showing a more abrupt break in the column density distribution \citep{2002ApJ...568L..71Z}.

\bsp	
\label{lastpage}
\end{document}